\documentclass[aps,pra,reprint,superscriptaddress]{revtex4-2}

\usepackage{graphicx}

\usepackage{color}
\usepackage{amsmath,amssymb}
\usepackage{bm}
\usepackage{upgreek}
\usepackage{xspace}

\usepackage[colorlinks,urlcolor=blue,citecolor=blue,linkcolor=blue]{hyperref}
\graphicspath{{figures/}}

\newcommand{\uvect}[1]{\hat{\boldsymbol{#1}}\xspace}

\begin{document}

\title{DC Quantum Magnetometry Below the Ramsey Limit}

\author{Alexander~A.~Wood}
\affiliation{School of Physics, University of Melbourne, Victoria 3010, Australia.}
\author{Alastair Stacey}
\affiliation{School of Science, STEM College, RMIT University, Melbourne, VIC 3001, Australia}
\author{Andy M. Martin}
\affiliation{School of Physics, University of Melbourne, Victoria 3010, Australia.}

\date{\today}

\begin{abstract}
We demonstrate quantum sensing of dc magnetic fields that exceeds the sensitivity of conventional $T_2^\ast$-limited dc magnetometry by more than an order of magnitude. We used nitrogen-vacancy centers in a diamond rotating at periods comparable to the spin coherence time, and characterize the dependence of magnetic sensitivity on measurement time and rotation speed. Our method up-converts only the dc field of interest and preserves the quantum coherence of the sensor. These results definitively improve the sensitivity of a quantum magnetometer to dc fields, an important and useful addition to the quantum sensing toolbox.     
\end{abstract}

\maketitle
Magnetometers based on the nitrogen-vacancy (NV) center in diamond~\cite{doherty_nitrogen-vacancy_2013, schirhagl_nitrogen-vacancy_2014} provide $\upmu$T\,- nT sensitivity for single centers at ambient temperature and mm-to-sub-$\upmu$m length scales, making them attractive resources for studying biomagnetism~\cite{gille_quantum_2021}, solid state systems~\cite{tetienne_quantum_2017, casola_probing_2018} and nanoscale NMR~\cite{mamin_nanoscale_2013} in challenging real-world sensing environments~\cite{fu_sensitive_2020}. Since many magnetic phenomena of importance in navigation and biomagnetism manifest as slowly varying or static magnetic fields, intense effort has been devoted in particular to improving dc sensitivity~\cite{barry_sensitivity_2020}, focusing on the diamond material~\cite{balasubramanian_ultralong_2009, herbschleb_ultra-long_2019}, photon collection efficiency~\cite{clevenson_broadband_2015}, quantum control sequences to eliminate decoherence~\cite{lange_controlling_2012, mamin_multipulse_2014, bauch_ultralong_2018} and more recently, the addition of ferrite flux-concentrators~\cite{fescenko_diamond_2020, zhang_diamond_2021}. Many approaches to improving the measurement signal are frustrated by a commensurate increase in noise, limiting the attainable sensitivity.   

To date, Ramsey-type interferometry~\cite{ramsey_molecular_1950} is the optimum dc measurement sequence~\cite{rondin_magnetometry_2014, barry_sensitivity_2020}, and the sensitivity of Ramsey magnetometry is limited by the ensemble dephasing time $T_2^\ast$, which reflects the magnitude of low frequency noise in the system. In diamond, impurities such as $^{13}$C~\cite{childress_coherent_2006} or paramagnetic nitrogen~\cite{bauch_decoherence_2020} are the dominant contributions to $T_2^\ast$, resulting in $T_2^\ast<1\,\upmu$s for readily-available CVD diamond samples. Additionally, $T_2^\ast$ varies considerably between diamond samples, and often significantly \emph{within} a single sample due to gradients or spatial variations of crystal strain or varied dopant density. This decoherence can be largely eliminated by employing time-reversal dynamical decoupling measurement schemes, such as Hahn spin-echo~\cite{hahn_spin_1950}, but at the cost of insensitivity to slowly varying or static dc fields~\cite{taylor_high-sensitivity_2008}. Alternative NV-magnetometry techniques~\cite{acosta_broadband_2010, jeske_laser_2016, wickenbrock_microwave-free_2016} that eschew conventional quantum sensing protocols have also been proposed and demonstrated, though to date yielded comparable sensitivities to standard methods.  

In this work, we use physical rotation to realize a significant improvement in the sensitivity of a dc quantum magnetometer. Mechanically rotating a diamond at rates comparable to the spin coherence time $T_2\sim 0.1-1$\,ms up-converts an external dc magnetic field to the rotation frequency, which is then detected using NV spin-echo magnetometry~\cite{wood_t_2-limited_2018}. The quantum sensing time is increased from $T_2^\ast$ up to $T_2$, with a potential $\sqrt{T_2/T_2^\ast}$ sensitivity gain, typically an order of magnitude~\cite{taylor_high-sensitivity_2008}. We demonstrate a 30-fold improvement in sensitivity over comparable Ramsey magnetometry. Furthermore, our demonstrated sensitivity exceeds, by a factor of 4.5, the theoretical shot-noise-limited, unity duty cycle sensitivity of $T_2^\ast$-limited Ramsey magnetometry with this diamond. 

\begin{figure*}
	\centering
		\includegraphics[width = \textwidth]{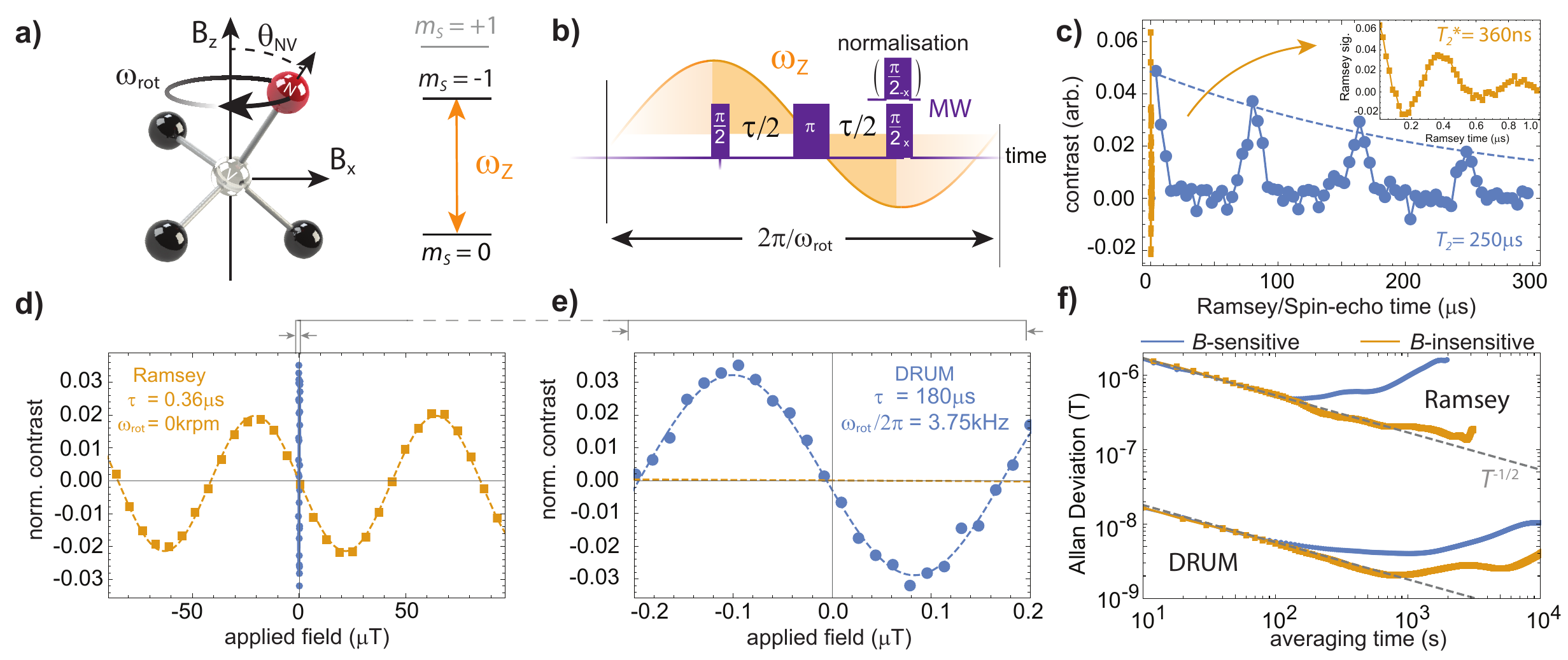}
	\caption{Diamond rotation up-conversion magnetometry (DRUM) schematic and key results. a) Schematic model of NV center, axis tilted from $z$ by $\theta_\text{NV} = 30.2^\circ$, rotating at $\omega_\text{rot}$ with external magnetic field components $B_z$ and $B_x$. b) Rotation of the diamond modulates the Zeeman shift in proportion to $B_x$, at a rate set by the rotation frequency. c) Typical stationary spin-echo (blue circles) and Ramsey (orange squares, detail inset) signals at $B_z = 2.3\,$mT for the diamond sample used in this work. d) Comparing magnetometry signals: varying an applied dc $B_x$ field (DRUM) or effective $z$-field (Ramsey) at the optimum sensing times traces out fringes. We used 10\,s and 30\,s measurement times per point, for Ramsey and DRUM, respectively. (e, zoom-in of d) The DRUM fringes are far faster, as considerably more phase accumulates within the $500\times$ longer sensing time. f) Allan deviation of DRUM and Ramsey when sensitive (blue) and insensitive (orange, microwaves detuned) to magnetic fields. Both measurements follow $T^{-1/2}$ scaling for a few hundred seconds, before drifts become dominant.}
	\label{fig:fig1}
\end{figure*}

Our method, depicted in Fig. \ref{fig:fig1}(a,b), is called diamond rotation up-conversion magnetometry (``DRUM") and was introduced in Ref.~\cite{wood_t_2-limited_2018}. The NV Hamiltonian in the presence of a magnetic field $\boldsymbol{B}$ is $H = D_\text{zfs} S_{z'}^2 + \gamma_e \boldsymbol{B}\cdot\boldsymbol{S}$, with $D_\text{zfs}/2\pi = 2870\,$MHz at room temperature, $\gamma_e/2\pi = 2.8\,$kHz\,$\upmu$T$^{-1}$ and $\boldsymbol{S} = (S_{x'}, S_{y'}, S_{z'})$ the vector of spin-1 Pauli matrices. We have ignored strain, electric fields and coupling to temperature. The quantization axis $z'$ is taken as the nitrogen-vacancy axis, lying along one of the four $[111]$ crystallographic axes of the crystal and therefore at an angle $\theta_\text{NV}$ to $z$. We use a single-crystal diamond containing an ensemble of NV centers, with a surface normal parallel to the rotation axis. A bias magnetic field is applied along the rotation axis $z$ to spectrally select one NV class and isolate $m_S$ transitions. The energies of the NV $m_S = 0, \pm1$ spin states for weak applied fields ($\hbar = 1$) are $\omega(m_S) \approx D_\text{zfs}+m_S \gamma_e B\cos(\theta_\text{NV})$, with $B = |\boldsymbol{B}|$. A weak magnetic field $B_x$ is applied along the lab-frame $x$-axis, making the Zeeman shift time-dependent during rotation at an angular frequency $\omega_\text{rot}$. Considering the $|m_S = 0\rangle-\leftrightarrow|m_S = -1\rangle$ transition, the time dependent component is 
\begin{equation}
\omega_{-1,0}(t) \approx \gamma_e B_x\sin\theta_\text{NV}\cos(\omega_\text{rot}t - \phi_0),
\label{eq:zeemant}
\end{equation}
with $\phi_0$ set by the initial orientation of the diamond, and adjusted to maximize sensitivity to either $x$ or $y$-oriented fields. The dc field in the lab frame is now effectively an ac field in the  NV frame, with amplitude $\delta B_x\sin\theta_\text{NV}$ and frequency $\omega_\text{rot}$. 

Our previous realization of DRUM used an NV tilt angle $\theta_\text{NV} =4^\circ$, resulting in up-conversion of only a small fraction of the dc field. In this work, we use a $\langle 110\rangle$-cut CVD-grown type IIa diamond with a natural abundance of $^{13}$C and approximately [N] = 1\,ppm, [NV] = 0.01\,ppm, mounted on an electric motor that can spin at up to 5.83\,kHz. We choose an NV orientation class making an angle of $30.2^\circ$ to the $z$-axis for our measurements, yielding $\sin\theta_\text{NV} = 0.5$, which is easily resolved and still sufficient to attain sensitivities exceeding that of optimized Ramsey sensing. Typical stationary spin-echo and Ramsey signals are shown in Fig. \ref{fig:fig1}(c) for $B_z$ = 2.3\,mT. The $^{13}$C spin bath is the dominant source of decoherence~\cite{barry_sensitivity_2020}, limiting $T_2^\ast$ to less than half a microsecond, but the sample exhibits a much longer $T_2 = 250(9)\,\upmu$s, with interferometric visibility restricted to revivals spaced at twice the $^{13}$C Larmor period~\cite{childress_coherent_2006}. Such a wide gulf separating $T_2$ and $T_2^\ast$ is not unusual, reflecting the much greater sensitivity of $T_2^\ast$ to the presence of magnetic impurities and sample-specific imperfections~\cite{bauch_ultralong_2018, bauch_decoherence_2020}.   

The remaining experimental details are similar to that described in Refs.~\cite{wood_quantum_2018, wood_t_2-limited_2018, wood_anisotropic_2021}, and further details are provided in the Supplementary Material. Briefly, a scanning confocal microscope optically polarizes (1\,mW 532\,nm) and reads (600-800\,nm, 9$\times10^6$ cts/s) the NV fluorescence. Microwave fields are applied along the $z$-axis with a coil antenna to ensure rotational symmetry. Further coils supply the $z$-oriented bias field and create the transverse dc test fields. A fast pulse generator controls the timing of laser and microwave pulses, and is triggered synchronously with the rotation. The laser is pulsed for $3\,\upmu$s to optically pump NVs to the $m_S = 0$ state, and the subsequent microwave spin-echo sequence is timed so that the $\pi$-pulse is applied at the zero-crossing of the up-converted field (Fig. \ref{fig:fig1}(b)), conferring maximum sensitivity. After the microwave pulses, optical readout is performed after a shuttling time so that the whole sequence takes one rotation period. A second sequence is applied back-to-back, with the final $\pi/2$-pulse phase shifted by $180^\circ$. The detected photoluminescence from each trace is then normalized and the difference computed to extract the contrast $\mathcal{S}$, which constitutes the DRUM signal.

The concept of rotational dc up-conversion should be applicable to a wide range of competing magnetometry architectures, including laser and absorption based readout schemes. The point of this paper is to demonstrate the sensitivity advantage of the DRUM technique over standard $T_2^\ast$-limited Ramsey magnetometry. Our priorities are therefore not to improve the ultimate sensitivity of the measurement, but rather ensure that the two techniques are compared on an equal basis. Using isotopically-purified diamonds with very high NV densities, wide-area collection optics and powerful excitation beams has been shown to be the most effective means to achieve high sensitivities~\cite{wolf_subpicotesla_2015}, and these techniques should be compatible with diamond rotation, with sufficient engineering application. The shorter coherence times intrinsic to these NV-dense configurations place additional requirements on rotation speed, and these will be discussed later in this work.

Due to strain and impurity inhomogeneity, $T_2^\ast$ exhibits considerable spatial variation\footnote{See Supplementary Material}. To assess the peak Ramsey sensitivity, we locate a region with a comparatively high $T_2^\ast = 360\,$ns and determine the sensitivity, $\delta B = \left(\frac{d\mathcal{S}}{dB}\right)^{-1} \sigma(\mathcal{S}) \sqrt{T}$, with $d\mathcal{S}/dB$ the mid-fringe signal slope, $T$ the total integration time, and $\sigma(\mathcal{S})$ the standard deviation of the Ramsey signal. For Ramsey, changing the microwave frequency is equivalent to varying a magnetic field exactly parallel to the NV axis. Ramsey fringes as a function of effective magnetic field are shown in Fig. \ref{fig:fig1}(d). The sensitivity of Ramsey magnetometry was found by detuning to the mid-fringe point and repeating the same 10\,s averaging interval 10 times. The best we achieved was a standard deviation of this data yielding $\delta B =0.86\upmu$T\,$\text{Hz}^{-1/2}$.

Next, we rotated the diamond at 3.75\,kHz and performed DRUM with a spin-echo time of $\tau = 180\,\upmu$s ($B_z = 0.7\,$mT), yielding fringes as an applied $x$-field is varied as shown in Fig.\ref{fig:fig1}(d,e), with each point averaged for 30\,s. We calculate the operational sensitivity to be $28\,\text{nT}\,\text{Hz}^{-1/2}$, about 30 times better than that of Ramsey in the same diamond sample. DRUM exhibits similar long-time averaging behavior to Ramsey, but towards a much lower minimum detectable field. Figure \ref{fig:fig1}(f) shows the Allan deviation~\footnote{A comprehensive description of Allan deviation as it applies to our work is provided in the Supplementary Material.} of DRUM and Ramsey as a function of averaging time. To assess the relative magnitudes of magnetic drifts and intrinsic noise in each technique, we measure while sensitive to magnetic fields, and then again with the microwaves detuned by 15\,MHz, so that only intrinsic measurement noise is present. Ramsey, sensitive to drifts in temperature, local strain as well as magnetic drifts (common also to DRUM) exhibits a significantly higher Allan deviation compared to DRUM. No amount of averaging time with Ramsey can exceed the performance of DRUM. This demonstration constitutes a significant achievement, showing that $T_2^\ast$ does not intrinsically limit dc magnetic sensitivity for a spin-based quantum magnetometer. In what follows, we describe how the optimum parameters were deduced, and how the ultimate sensitivity depends on parameters we can control.

The shot-noise limited dc sensitivity of the DRUM measurement is given by~\footnote{See Supplementary Material}
\begin{equation}
\delta B = \frac{\pi e^{\left(\frac{\tau}{T_2}\right)^n}}{4 C \gamma_e \sin\theta_\text{NV}\sin\left(\frac{\pi\tau}{2t_\text{rot}}\right)}\frac{1}{\sqrt{t_\text{rot}}},
\label{eq:sensitivity}
\end{equation}
with $n\approx 3$ reflecting sample-specific decoherence processes, $C$ the readout efficiency~\cite{degen_quantum_2017} and $t_\text{rot}$ the rotation period. Extracting the optimum performance for DRUM amounts to increasing the slope $d\mathcal{S}/dB$, which depends on the rotation speed and spin-echo measurement time, and increasing the measurement signal-to-noise, largely via well-understood means~\cite{barry_sensitivity_2020} that benefit both Ramsey and DRUM.

The rotation speed can be chosen to optimize sensitivity. The intrinsic amplitude of the up-converted field is set by the lab-frame dc field, but the phase accumulated by the NV centers depends on the rotation speed. Faster rotation speeds enable the same integration time to sample a greater fraction of the modulated field, and more measurements are possible within a given averaging time, increasing the number of photons collected. However, without recourse to higher-order dynamical decoupling sequences, which supplant spin-echo as the optimum measurement only when $T_2\gg t_\text{rot}$, increasing the rotation speed eventually reduces the accumulated phase due to the smaller integrated area under $\omega_{-1,0}(t)$. 

Maximizing the slope also requires minimizing decoherence sources. For a natural abundance diamond, the NV-$^{13}$C bath interaction depends on the strength~\cite{zhao_decoherence_2012, hall_analytic_2014} and direction~\cite{stanwix_coherence_2010, wood_anisotropic_2021} of the magnetic field. Consequently, the $z$-bias field tunes the spin-echo time $\tau$, and must be set so that $\tau = n_c / (B\times 10.71\,\text{kHz\,mT}^{-1}\pm\omega_\text{rot}/2\pi)$, $n_c\in\{1, 2, ...\}$. Stronger $z$-fields result in faster decoherence due to the anisotropic NV-$^\text{13}$C hyperfine coupling~\cite{stanwix_coherence_2010,wood_anisotropic_2021}. We therefore minimize the requisite $z$-field by ensuring just the \emph{first} $^{13}$C revival coincides with the desired measurement time, and choose the rotation direction so that the induced magnetic pseudo-field~\cite{wood_magnetic_2017} adds to the $z$-bias field, \emph{i.e.} $\omega_\text{rot} >0$.

\begin{figure}[t!]
	\centering
		\includegraphics[width = \columnwidth]{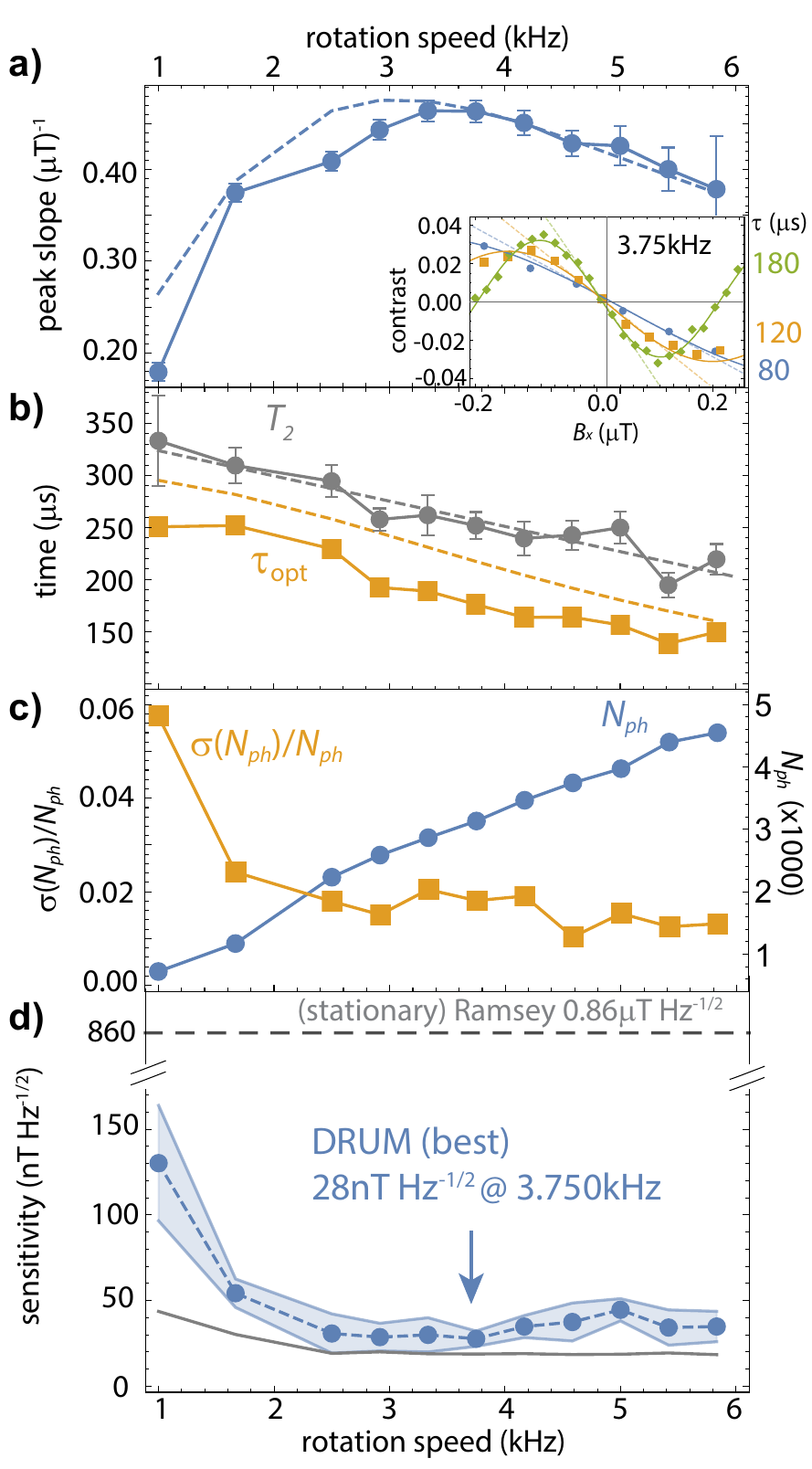}
	\caption{Optimization of DRUM. (a) Peak slope $d\mathcal{S}/dB$ as a function of r$\omega_\text{rot}$, deduced from fits of Eq. \ref{eq:slope} to slope-vs-$\tau$ data gathered for DRUM fringes at different rotation speeds (inset, 3.75\,kHz) (b) Measured $T_2$ (gray dashed line is linear fit) and spin-echo time $\tau_\text{opt}$ where the peak slope occurs: decoherence results in $\tau_\text{opt} < T_2$. Dashed lines in (a) and (b) correspond to theoretical predictions from measured $T_2$ data. (c) Increasing photon number per measurement bin (blue circles), and reducing measurement noise (orange squares). (d) Sensitivity $\delta B$ as a function of rotation speed. Shaded regions denote maximum and minimum ranges from 5 repetitions.}
	\label{fig:fig2}
\end{figure}

The slope $d\mathcal{S}/dB$ is given by 
\begin{equation}
\frac{d\mathcal{S}}{dB} = \frac{4A}{\omega_\text{rot}}e^{-\left(\frac{\tau}{T_2}\right)^n}\gamma_e \sin\theta_\text{NV}\sin\left(\frac{\omega_\text{rot}\tau}{4}\right).
\label{eq:slope}
\end{equation}
For comparison with data, we leave $A$ and $T_2$ as free parameters, possibly dependent on rotation speed, and fix $n = 3$, which best describes our observed relaxation and is consistent with other experiments~\cite{stanwix_coherence_2010, hall_analytic_2014}. We measured the mid-fringe slope of DRUM fringes as a function of $\tau$ and $\omega_\text{rot}$. Fig.~\ref{fig:fig2}(a, inset) shows DRUM fringes as the spin-echo time is varied for $\omega_\text{rot}/2\pi = 3.75\,$kHz, and the slope extracted from fits of Eq. \ref{eq:slope} to such data is plotted versus $\tau$ for rotation speeds from $1-6\,$kHz (Fig.~\ref{fig:fig2}(a)). We find that as the rotation speed is increased, an optimal sensing time appears around 3 to 4\,kHz. 

We also monitored the coherence time $T_2$ and time at which the peak slope occurs $\tau_\text{opt}$ as a function of rotation speed (Fig. \ref{fig:fig2}(b)). The measured $T_2$ is higher than that in Fig. \ref{fig:fig1}(c) due to the overall lower magnetic field strengths, and drops slightly as a function of rotation speed, which we believe is due to imperfectly canceled $y$-fields or the test field itself inducing weak anisotropy in the NV-$^{13}$C interaction~\footnote{See Supplementary Material}. Numerically maximizing Eq. \ref{eq:slope} for the measured $T_2$ yields the dashed theoretical predictions in Fig.~\ref{fig:fig2}(a, b), confirming the key role $T_2$ plays. While the theory accurately reproduces the peak slope, the optimum time $\tau_\text{opt}$ where the slope is maximized is lower than predicted; we attribute this to the particular choice of $n = 3$ in Eq. \ref{eq:slope}.

The variation in slope is tempered by increasing photon collection rates (and hence reduction in shot noise) due to increased duty cycle at higher rotation speeds, as shown in Fig. \ref{fig:fig2}(c). This leads to an almost flat dependence on sensitivity with rotation speed, as shown in Fig. \ref{fig:fig2}(d). DRUM operates close to the shot-noise predicted sensitivity limit, 30 times better than the Ramsey sensitivity. This is actually greater than $\sqrt{T_2/T_2^\ast} = 26$ due to the $10\%$ duty cycle of Ramsey measurements (including laser preparation and readout), and the trigonometric factor and integration for less than a full period in DRUM. However, computing the idealized Ramsey sensitivity with unity duty cycle yields $122$\,nT\,Hz$^{-1/2}$, still a factor of 4.5 worse than experimentally demonstrated DRUM with $\tau < t_\text{rot}$ and $\theta_\text{NV} = 30^\circ$.

\emph{Discussion.} We have shown in this work that dc up-conversion magnetometry with a rotating diamond can significantly exceed the sensitivity of conventional Ramsey magnetometry. Unlike previous proposals~\cite{ajoy_DC_2016}, our scheme up-converts just the magnetic fields of interest and not the deleterious noise that limits quantum coherence, thus definitively improving sensitivity to dc fields. We also retain the vector sensitivity of the NV center, with a timing adjustment to the synchronization allowing for exclusive $y$-field detection. In principle, our scheme is equally applicable to any quantum system where the coupling between qubit and parameter of interest can be modulated in time, and this paper shows that sensitivity exceeding the $T_2^\ast$ limit is thus possible. 

For up-conversion to be worthwhile, $T_2\gg T_2^\ast$ and $t_\text{rot} \sim T_2$. More practically, however, the $\sqrt{\omega_\text{rot}}$ scaling of sensitivity in Eq. \ref{eq:sensitivity} requires a commensurate increase in $C$, \emph{i.e.} by increasing NV density, which in turn reduces $T_2$~\cite{bauch_decoherence_2020}. However, with the simplified assumption that $\tau = t_\text{rot} = T_2$ and $\theta_\text{NV} = 90^\circ$, DRUM can confer a sensitivity gain over Ramsey in any diamond sample of $\delta B_\text{DRUM}/\delta B_R  = \frac{\pi}{4}\sqrt{T_2^\ast/T_2}$. For DRUM to have a factor of 10 better sensitivity than unity-duty-cycle Ramsey, $T_2 =62\,T_2^\ast$.

Mechanical rotation can be challenging, though not impossible, to achieve for the short $T_2\sim 5-10\,\upmu$s exhibited by NV-dense diamond samples. Commercial NMR magic-angle spinning devices can achieve rotation rates of mm$^3$-scale samples now up to 150\,kHz~\cite{schledorn_protein_2020}, and demonstrations of ultrafast rotation of optically-trapped microscale structures~\cite{reimann_GHz_2018, ahn_ultrasensitive_2020} yield GHz rotation frequencies. Since rapid libration can be substituted for rotation, alternative approaches could potentially leverage fast piezoelectric tip-tilt transducers~\cite{csencsics_fast_2019} for larger samples or micro-to-nano structures for smaller length scales, for instance in optically or electrically-trapped~\cite{perdriat_spin-mechanics_2021} micro-to-nano diamonds. Modulation can also be achieved by position displacement in a spatially-varying field, and recent work has demonstrated up-conversion of dc fields to ac using scanning single NV magnetometry by rapidly modulating the distance between tip and sample~\cite{huxter_scanning_2022}. Another option could be modulation, \emph{i.e.} via position displacement, of the sensitivity enhancement conferred by ferrite flux concentrators~\cite{fescenko_diamond_2020}.   

In conclusion, we have demonstrated that $T_2$-limited dc magnetometry can exceed the sensitivity of $T_2^\ast$-limited Ramsey magnetometry for diamond-based quantum magnetometers. We anticipate our work will stimulate other approaches, not necessarily based on sample rotation, to combine up-conversion with existing schemes to improve magnetic sensitivity. Augmented with fast rotation, improvements such as engineered diamonds with higher NV densities and larger optical excitation and collection areas may be sufficient to achieve the much sought-after fT\,Hz$^{-1/2}$ regime of dc magnetic sensitivity with room-temperature, microscale diamond sensors, where applications such as magnetoencephalography in unshielded environments become possible~\cite{boto_moving_2018}.

This work was supported by the Australian Research Council (DE210101093, DE190100336). We thank R. E Scholten for insightful discussions and a careful review of the manuscript. The authors (AAW, AS, AMM) are inventors on a United States Patent, App. 16/533,167 which is based on this work.

\newpage
\widetext
\begin{center}
\textbf{\large DC Quantum Magnetometry Below the Ramsey Limit - SUPPLEMENTARY INFORMATION}
\end{center}

\setcounter{equation}{0}
\setcounter{figure}{0}
\setcounter{table}{0}
\setcounter{page}{1}
\makeatletter

\renewcommand{\thesection}{S\arabic{section}}
\renewcommand{\thefigure}{S\arabic{figure}}
\renewcommand*{\citenumfont}[1]{S#1}
\renewcommand*{\bibnumfmt}[1]{[S#1]}

\section{Experimental setup}

The basic configuration of the experiment is the same as reported in Ref. \cite{wood_t_2-limited_2018s}, with several differences and improvements. The experimental configuration is depicted in Fig. \ref{fig:figs1} and a description of the key elements follows.

\subsection{Confocal Microscope} The optical illumination and photon collection is performed by a scanning confocal microscope, mounted in a top-down configuration as shown in Fig. \ref{fig:figs1}. A 50x magnification, $NA = 0.6$ Nikon SLWD objective with an $11\,$mm working distance is mounted to a $xyz$-piezoelectric nanopositioner stage (Physik Instrumente P-545.R7 series) with $200\,\upmu$m travel range in each direction. About $1\,$mW of 532\,nm laser light (Laser Quantum Opus) is input to the back aperture of the objective. We collect $8-10\times10^6\,$ photons/s on a single-photon counting module (SPCM, Excelitas SPCM-AQRH-14) from the diamond sample after coupling into a $50\,\upmu$m-diameter multi-mode optical fiber with a 250\,mm achromatic doublet lens. Photon counts and arrival times are registered on a fast counting module (FAST ComTech MCS6A) and analyzed on a dedicated computer.

\subsection{Diamond Sample and motor} We used two similar diamond samples and two ideally identical motors in this work to better assess the implementation variability of DRUM. Each diamond sample is an optical-grade ($[N]<1\,$ppm) type IIa single-crystal diamond plate from Delaware Diamond Knives (DDK), $3\times1.2\times0.2$\,mm and $\langle110\rangle$-cut wide face with a $\sim4^\circ$ polishing miscut. Each sample has a reasonably uniform distribution of NV centers throughout though with notable bright stripes of NV fluorescence, and we undertake no pre-treatment other than sample cleaning. The diamond samples are bonded with UV-curing epoxy (Norland Optical Adhesive Type 63) to the end of a high-speed electric motor (Celeroton CM-2-500) driven by a three-phase converter (Celeroton CC-75-500 with E08 synchronization extension board). The motors can be ran at a maximum speed of 500,000\,rpm in either direction, though for mechanical longevity we restrict speeds below 350,000\,rpm. 

Individual motors exhibit some variation in stability for different rotation speeds (occasionally, audibly so), and we observe a run-in effect whereby sustained operation at a single speed (\emph{i.e.} for months at a time) makes the motor unstable at other speeds, larger or smaller. We experienced this effect with one motor and replaced it with a different motor and diamond that had not been run-in, and this data was presented in the paper. We also examined the effects of different rotation directions on DRUM, as discussed later in Section \ref{sec:rot}. 

\begin{figure}[t]
	\centering
		\includegraphics[width = \columnwidth]{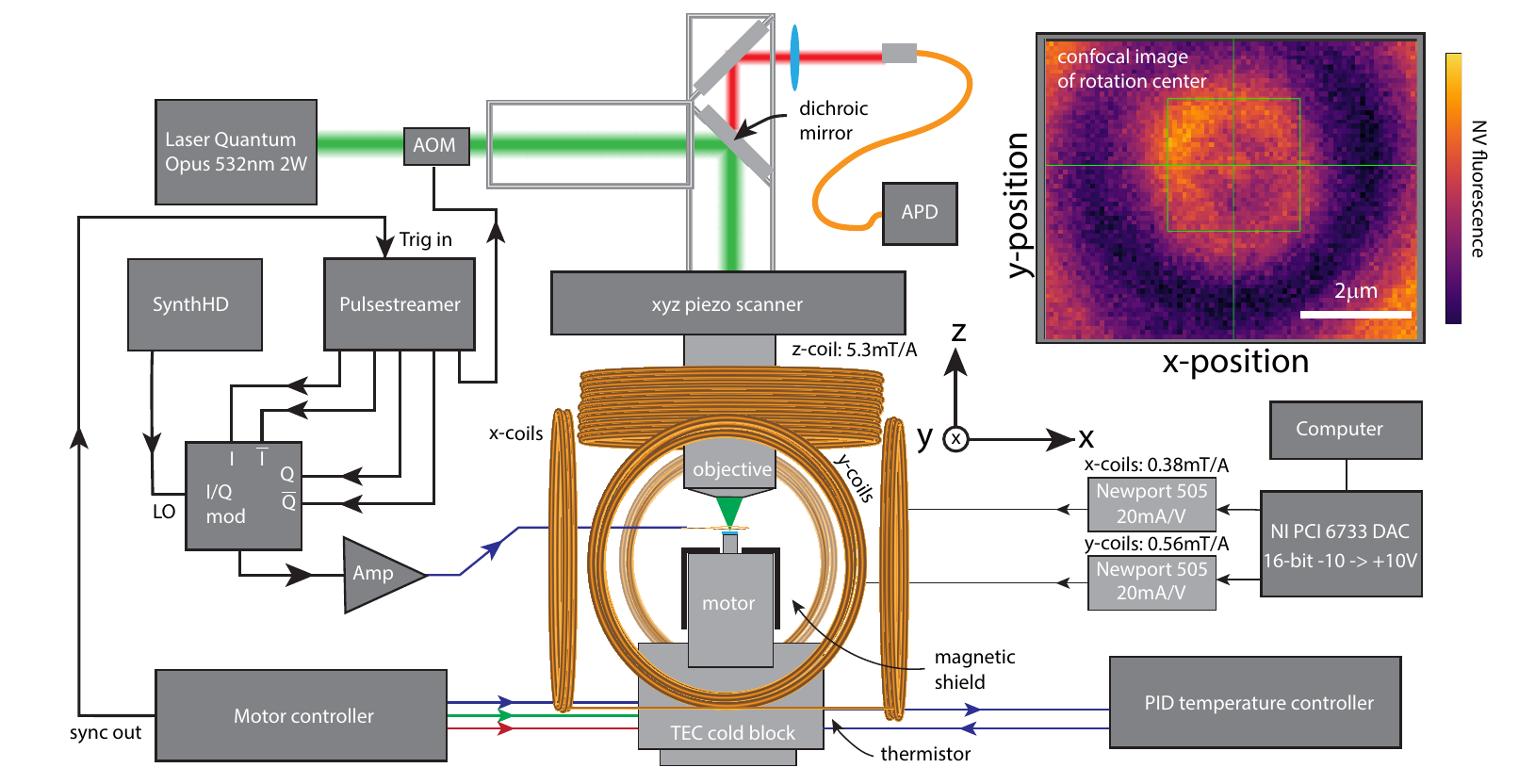}
	\caption{Block diagram of experimental setup and magnetic field geometry. Inset image shows scanning confocal image near rotation center, crosshair denotes point of symmetry. }
	\label{fig:figs1}
\end{figure}

We determine the `rotation center', \emph{i.e.} the $r = 0$ point about which the diamond rotates by performing unsynchronized $x-y$ scanning confocal microscopy and identifying the characteristic patterns of circular motion. Inhomogeneities are blurred into circular patterns centered at $r = 0$, and we position the microscope objective at this point to reduce the effects of off-axis rotation, as shown in Fig. \ref{fig:figs1}. The motor is securely recessed into a thick block of aluminium, and a commercial PID controller stabilizes the motor temperature to better than 0.01\,K. A 1.0\,mm thick mu-metal shield covers the exposed part of the motor and suppresses magnetic interference from the rotating pole piece.
          
\subsection{Microwave generation and control} We use a Windfreak Technologies SynthHD PLL-based microwave generator as a local oscillator (LO) feeding an I/Q modulator (Texas Instruments TRF37T05EVM IQ modulator evaluation board). We drive the $I, \bar{I}, Q, \bar{Q}$ inputs of the I/Q modulator using TTL pulses from a Swabian Instruments Pulsestreamer 8/2, which allows for fast, almost distortionless pulsing with phase control: a single TTL pulse into one of the I/Q channels yields the microwave pulse phases $\left[I, \bar{I}, Q, \bar{Q}\right] = \left[0, 180^\circ, 90^\circ, 270^\circ\right]$, and we observe LO suppression comparable to a commercial vector signal generator. The phase control, while limited is still useful: we use $180^\circ$ pulses, \emph{i.e.} a digital high into the $\bar{I}$ channel to normalize the spin-echo (or Ramsey) signals. The Pulsestreamer is hardware-triggered by a digital signal from a delay generator synchronized to the rotation of the motor.

Microwave fields are applied to the NV centers using a 2\,mm diameter, 5-turn coil fabricated from $150\,\upmu$m enamel-insulated copper wire located about $200\,\upmu$m above the surface of the diamond. We drive the coil with a 50\,W microwave amplifier (Microwave Amps AM6-1-3.5-47-47). As a result of the coil geometry and standoff distance, we achieve only modest Rabi frequencies, ranging from $6.24\,$MHz at a frequency of $2754\,$MHz ($B_z = 4.8\,$mT) to 2.45\,MHz at  2845\,MHz (1.0\,mT). A coil-type geometry is required to ensure a rotationally-symmetric mw Rabi frequency, and we carefully position the coil so that the Rabi frequency varies as little as possible as the diamond is stepped through fixed rotation angles (typically better than $50\,$kHz across a full period). 

\subsection{Magnetic fields} The $z$-oriented dc magnetic field for spectrally selecting NV orientation classes and tuning the $^{13}$C revival time is provided by a single coil coaxial with the laser, with a characteristic calibration of $5.3\,$mT/A. A programmable switched-mode power supply (HMP2030) provides up to 1\,A to drive the coil. Two pairs of coils enclosing the system provide $x$- and $y$-oriented magnetic fields, which are used cancel non-axial components of the $z$-applied field, and provide the test magnetic fields used in our experiments. The test fields are defined as the added current above that required to cancel the $z$-coil misalignment, \emph{i.e.} $B_x = i_x B_x' + B_{x0}(i_z)$ , $B_y = i_y B_y' + B_{y0}(i_z)$, with $i_{x, y, z}$ the $x$, $y$ or $z$ coil currents, $B_{x0}, B_{y0}$ the `nulling' fields, which depend on the $z$-coil current $i_z$, $B_x'$ = 0.38\,mT/A and $B_y'$ = 0.56\,mT/A. We determined these calibration factors using NV vector magnetometry and comparing against diagonalization of the NV Hamiltonian, the process is described fully in the Supplemental Material of Ref. \cite{wood_t_2-limited_2018s}. The $x$ and $y$-fields are driven by separate precision laser diode current drivers (Newport Model 505) which ensures low-noise, highly stable test magnetic fields. We drive the 20\,mA/V analog voltage input of the current drivers with a 16-bit DAC, yielding 15\,$\upmu$A current resolution.

\subsection{Improvements over previous experiment} Most improvements over the previous demonstration are easy to quantify directly, such as the increased NV tilt angle $\theta_\text{NV} = 30^\circ$, yielding a trigonometric factor $7\times$ the previous experiment. We also work with higher laser powers, a slightly higher NV density, $\sim2\times$ larger optical preparation and better photon collection, yielding $3$ times higher photon counts than Ref. \cite{wood_t_2-limited_2018s}, enabling us to reduce the averaging time for DRUM by a factor of 10. These same benefits (excepting the trigonometric factor) also apply to Ramsey, and we achieved a $4\times$ improvement in Ramsey over the previous measurement as well. The $T_2$ of the diamond we used in this work enables longer measurement times, $\tau = 180\,\upmu$s (and up to $200\,\upmu$s) vs $124\,\upmu$s previously.   

Other improvements are less easy to explicitly quantify, though definitively beneficial. In this work, we were able to run two DRUM sequences back-to-back in consecutive periods, alternating the readout $\pi/2$-pulse phase between 0 and $180^\circ$, which provides common-mode rejection of magnetic noise, notably that arising from AC mains interference. Furthermore, we optimized the pulse timing to eliminate trigger skipping, which randomly added dead time and occurs when the measurement pulse sequences is very near to the rotation period of the diamond. Neither of these issues applied to Ramsey magnetometry in either this or the previous realisation. Implementing ultra-stable current drivers for the shim and test fields is also highly beneficial, the resolution alone of current drivers used in the current work being 25 times finer than the previous case.

\subsection{General} We use the `Qudi' experimental control software \cite{binder_qudi_2017s} for managing computer control of the experiment. We note that the current version of the experiment exhibits remarkable magnetic stability, as depicted in Fig. 1(f) of the main text, and elaborated on further when we discuss Allan deviation in Sec. \ref{sec:adev}. We are able to run the motor at very high speeds (5\,kHz) for several days continuously, for periods of several months without noticeable degradation of the rotation. We note that such sustained operation at a fixed rotation speed appears to adversely affect the motor's stability at other speeds, the `run-in' effect mentioned previously. Given the relatively flat scaling of DRUM sensitivity, run-in poses an issue only for the characterisation measurements we performed. For general operation, we do not anticipate a case where varied rotation speed is essential.     

\section{Ramsey Variation}

\begin{figure}
	\centering
		\includegraphics[width = \columnwidth]{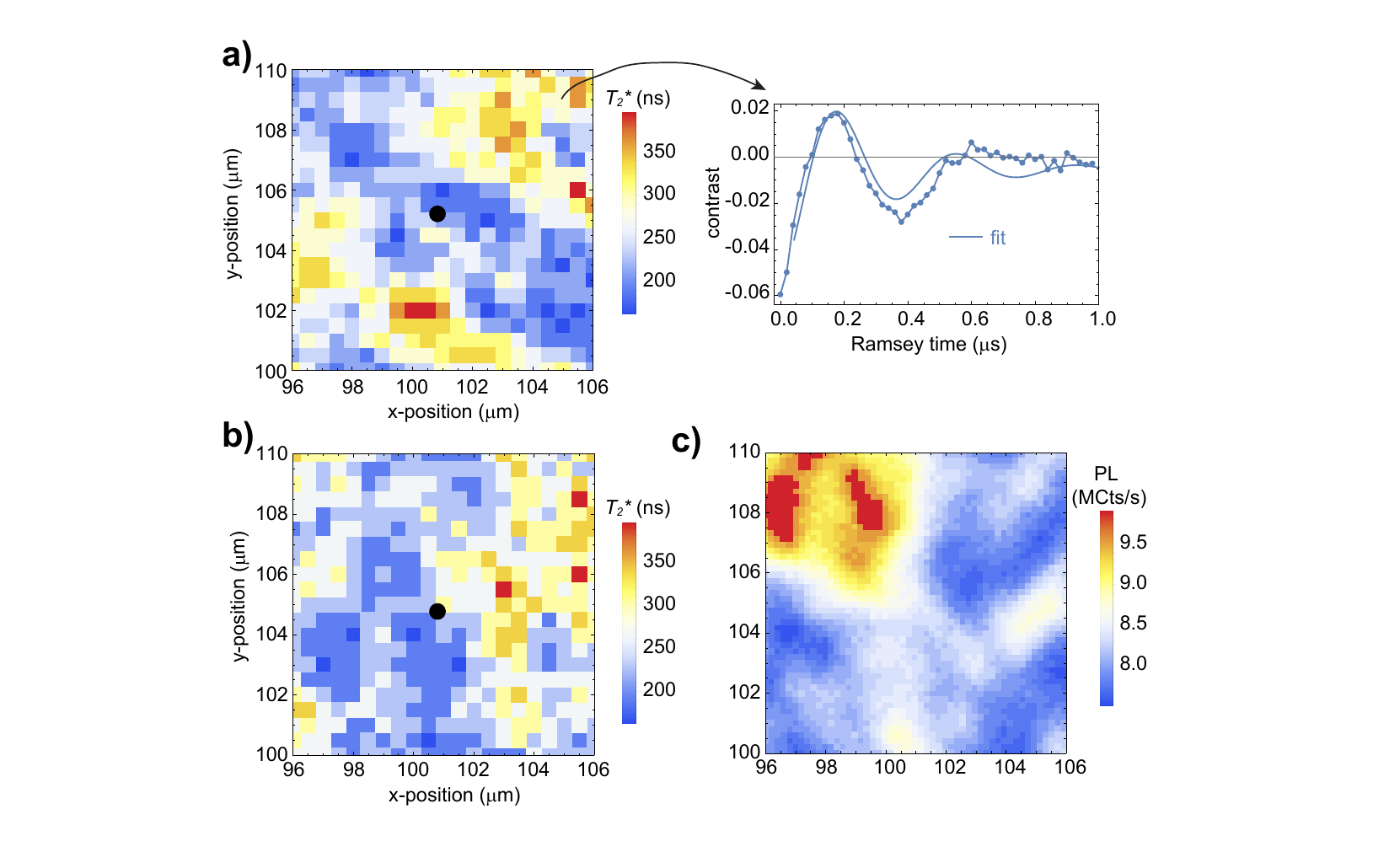}
	\caption{Spatial variation of Ramsey $T_2^\ast$. (a) We measured Ramsey fringes (right) as a function of confocal $x,y$ position in a 10\,$\upmu$m square. Fitting to the fringes yields $T_2^\ast$, which varies considerably over the region of interest (left). Black dot denotes approximate position of rotation center. (b) Rotating the diamond clockwise by 45$^\circ$ and repeating the measurement shows the spatial variation to vary with rotation angle, confirming it is predominantly intrinsic to the diamond. (c) While spatially varying crystal strain is the principal culprit for $T_2^\ast$ variation, we observe a rough correlation between NV PL (at $0^\circ$ motor park angle, a proxy for nitrogen density) and the variability in Ramsey coherence time in (a), note in particular the large bright patch in the top left corner corresponds to an equally sized region of low $T_2^\ast$. }
	\label{fig:figs2}
\end{figure}

A characteristic feature of both samples we investigated is strong spatial variation of $T_2^\ast$ over $\upmu$m-scale regions. We measured time-domain Ramsey signals as a function of confocal $x-y$ position in a (10$\,\upmu$m)$^2$ region, as shown in Fig. \ref{fig:figs2}(a). The $T_2^\ast$ extracted from damped sinusoidal fits varies between $160\,$ns and $390\,$ns over a few micrometers. This variation is attributed to local variation in crystal strain, as reported in Ref. \cite{bauch_ultralong_2018s} and possibly nitrogen density: we note a correlation between low values of $T_2^\ast$ and high NV fluorescence, as shown in Fig. \ref{fig:figs2}(c). Notably, this spatial pattern rotates when we rotate the diamond by `parking' the motor at different angles, as shown in Fig~\ref{fig:figs2}(b). This final observation confirms the origin of the inhomogeneity as inside the diamond. We also note (data not shown) that $T_2$ extracted from spin-echo measurements shows essentially no spatial variation. 

\section{Allan deviation}\label{sec:adev}

Fig. 1(f) of the main text depicts the long-term stability and averaging characteristics of DRUM, encapsulated in the \emph{Allan deviation}. The Allan deviation is primarily used a measure of precision oscillator and inertial sensor stability to date, has featured in several recent works in NV precision sensing, eg. Refs~\cite{wolf_subpicotesla_2015s, michl_robust_2019s, fescenko_diamond_2020s}. While much of the existing body of literature concerns application to precision time standards, the key features of Allan deviation can be ported to magnetometers, though not without some clarification to prevent misinterpretation. 

A key difference is in the intended use of the measure. For a clock, the variation of time is assumed to be well known, indeed immutable, and variability in an oscillator's measurement of time assumed to arise from noise and imperfections. A magnetometer is fundamentally different, in that the device itself is subject to imperfections and noise, but the quantity being measured, \emph{i.e.} the magnetic field, does \emph{not} follow a well-defined variation. For example, the NV zero-field splitting can drift as a result of temperature variations, a `device' defect, but the magnetic environment may also drift and fluctuate, an environmental issue unrelated to the performance of the magnetometer itself. This being the case, a magnetometer with low Allan deviation that follows a particular scaling is fairly meaningless in isolation and requires comparison with some measure of either the device's intrinsic defects or that of the environment. Examples are operation in a carefully controlled or magnetically shielded environment, or, operation without sensitivity to magnetic fields. The former is ideal though difficult to achieve, the latter is simple though somewhat less informative.

Allan deviation is fundamentally a measure of oscillator stability. The oscillator in our case is the NV two-level splitting, and we measure the relative phase between this and an applied microwave field in any interferometry experiment. We assume that the microwave field is essentially perfect on the timescale of our experiments, so that the `noisy' oscillator is the NV itself. In our experiment, we set DRUM to the mid-fringe sensing point with an appropriate applied $x$-field. We then run the same measurement continuously for several hours or more. The fast counting hardware accumulates a photoluminescence trace $S$ continuously, and we compute the NV contrast using only the photons counted in between polling times, typically 10\,s, \emph{i.e.}
\begin{equation}
S_n = S(t_n) - S(t_n + t_\text{poll}).
\end{equation}
The time series $\{S_n\}$ is then converted from NV fluorescent contrast to frequency perturbations using 
\begin{equation}
\omega_n = \gamma\left(\frac{d\mathcal{S}}{dB}\right)^{-1} S_n.
\end{equation}    
Assuming the deviation of the NV contrast is constrained to the linear region of $\mathcal{S}$, which it is in our experiments, there is no need for phase unwrapping, and the total phase $\phi_n$ of the NV as a function of time can be determined by the cumulative sum of $\omega_n$ multiplied by the polling time interval $\delta T$
\begin{equation}
\phi_n = \sum_{k = 1}^n \delta T \omega_k.
\label{eq:adph}
\end{equation}
This list of phase samples as a function of measurement time is what the Allan deviation assesses the stability of. The Allan deviation at an averaging time of $m\delta T$ for $N$ samples of total phase $\phi_n$ using the overlapped $T$ estimator is 
\begin{equation}
\sigma_A^2(m\delta T) = \frac{1}{2(m\delta T)^2 (N - 2m)}\sum_{i=0}^{N-2m-1}\left(\phi_{i+2m}-2\phi_{i+n}+\phi_i\right)^2.
\label{eq:adev}
\end{equation}

\begin{figure}
	\centering
		\includegraphics[width = \columnwidth]{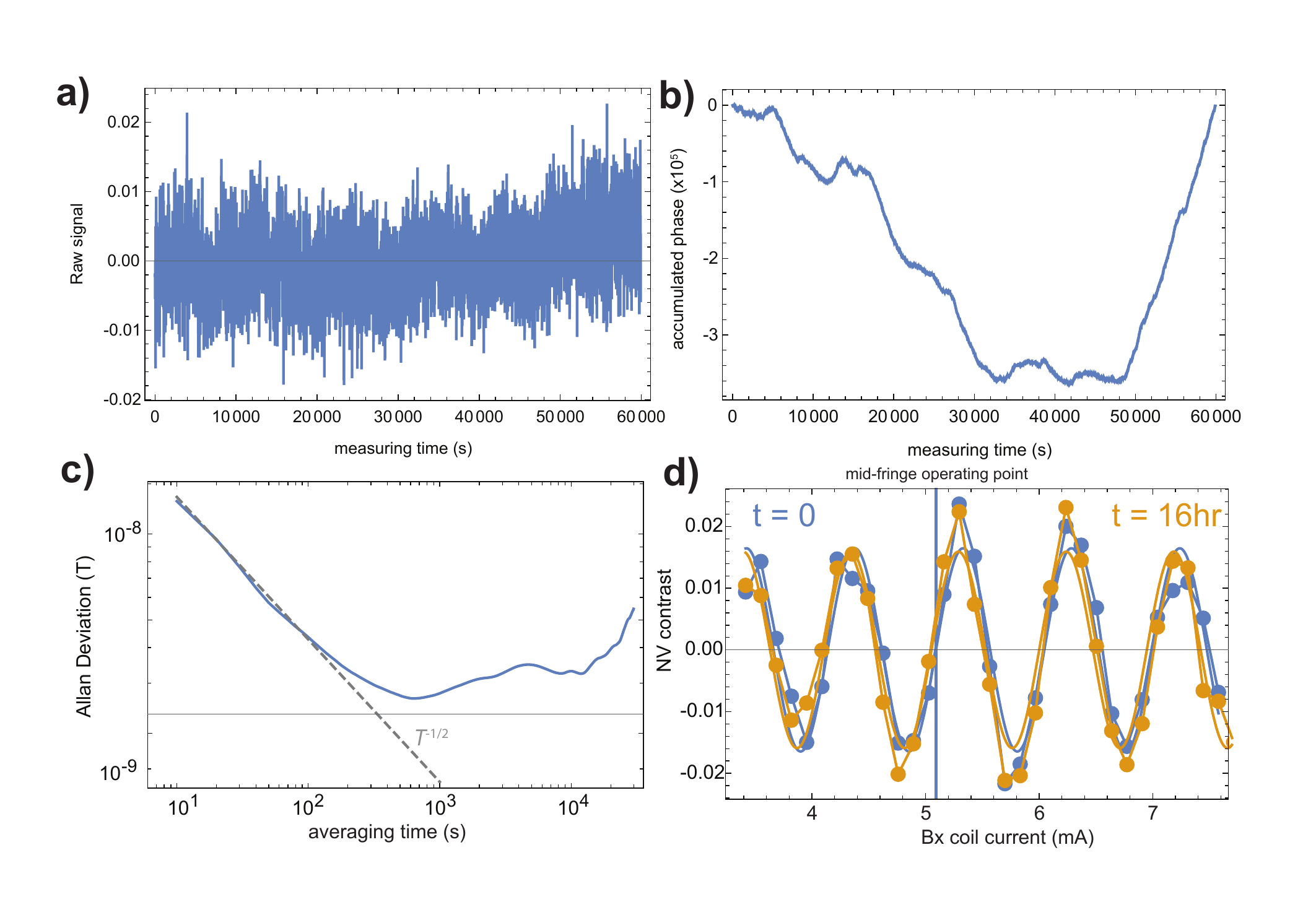}
	\caption{Allan deviation measurement of oscillator stability. (a) Raw NV contrast at the mid-fringe point, logged every 10\,s. (b) accumulated total phase, according to Eq. \ref{eq:adph}. (c) Calculated overlapping Allan deviation as a function of averaging time. The $1/\sqrt{T}$ scaling indicates white frequency noise, which originates from photon shot noise. (d) DRUM fringes (traced here as a function of bare $B_x$-coil current) before and after a 16\,hr run, showing minimal phase drift.}
	\label{fig:figsa}
\end{figure}

Figure \ref{fig:figsa} shows the raw NV contrast as a function of time (sampled every 10\,s), the corresponding total phase as a function of time and finally the Allan deviation. We confirm that our measurement stays `on the same fringe' by measuring DRUM fringes at the beginning and end of the Allan deviation data run, as shown in Figure \ref{fig:figsa}(d). The amplitude modulation we observe here is due to the long microwave pulses used, typically $t_\pi  = 200\,$ns.

So, what does the Allan deviation \emph{mean} for a magnetometer? In the absence of a power spectral density plot, the Allan deviation allows some information about the dominant sources of noise to be deduced. The $1/\sqrt{T}$ scaling of $\sigma_A$ we observe for averaging times up to 500\,s is consistent with white frequency noise, that is, random fluctuations in the measured oscillator frequency consistent with errors originating from photon shot noise. Beyond this time, the Allan deviation inflects and then increases as $\sqrt{T}$, consistent with a random walk (brown noise) and slow drift. 

Considering the data shown in the main text (Fig. 1(f)), the Allan deviation when insensitive to magnetic fields (i.e. microwaves detuned) continues to decrease as $1/\sqrt{T}$ for Ramsey, but shows more complicated behaviour (when $B$-sensitive \emph{and} $B$-insensitive) for DRUM. That the same features are present on the magnetometer signal when sensitive \emph{and} insensitive to magnetic fields says that the noise affects the NV-spin state readout. We attribute this primarily to drift of the confocal focus away from the rotation center, which if severe ($>2\,\upmu$m, the confocal spot size) can affect optical state preparation and readout due to timing errors and motor jitter. This drift primarily occurs due to lab temperature variations of order $1-2^\circ$C throughout long measurement runs. Drift of the confocal spot also introduces slow magnetic drifts, since magnetic field gradients (which we measure using DRUM by displacing from the rotation center) are present, resulting in a phase shift in DRUM fringes as the confocal spot is displaced.

\section{Peak sensing time and vector sensitivity}
\begin{figure}
	\centering
		\includegraphics[width = \columnwidth]{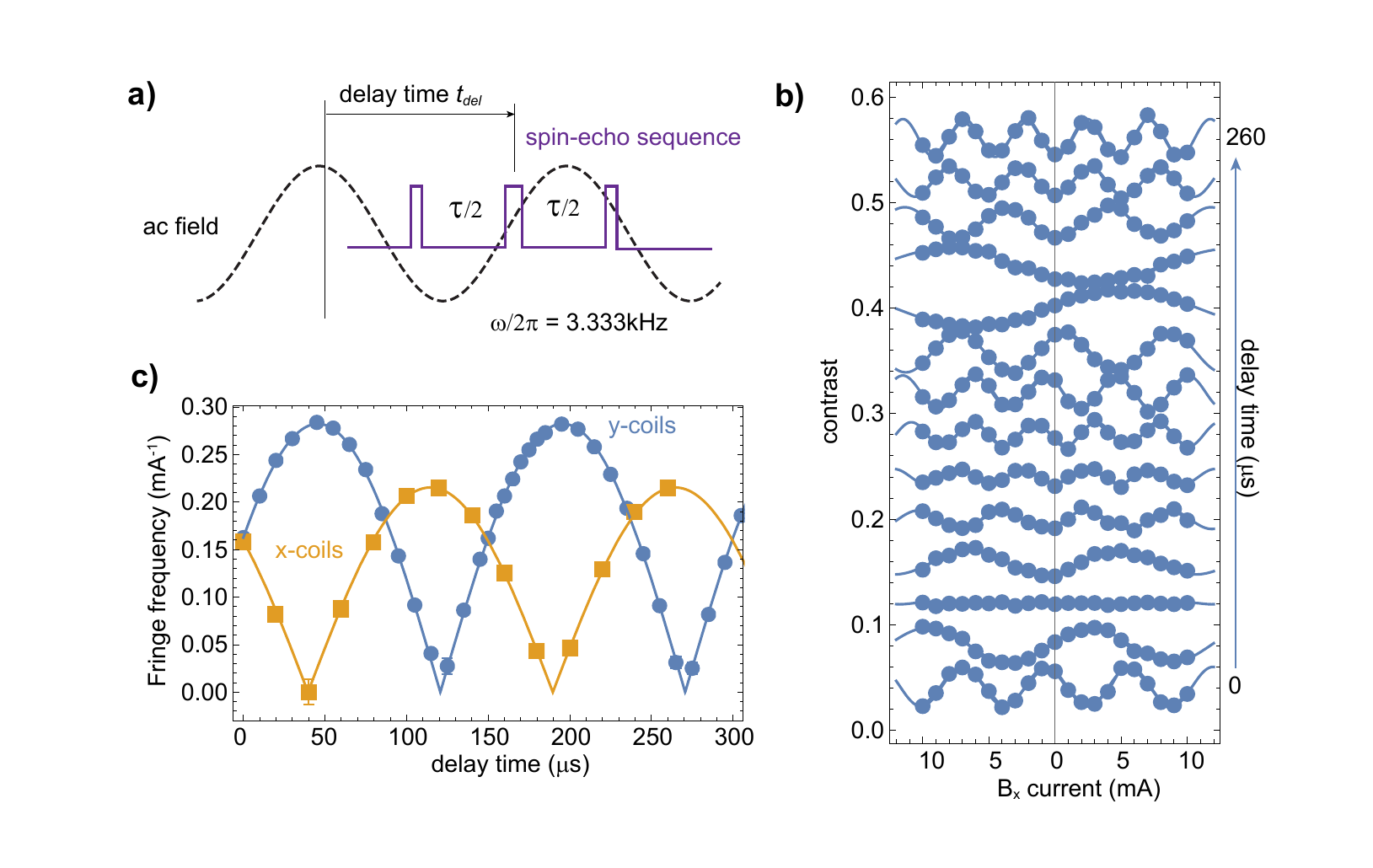}
	\caption{Optimisation of DRUM fringe frequency and vector sensitivity. (a) Schematic of procedure; we conduct a spin-echo measurement with $\tau = 70\,\upmu$s at $\omega_\text{rot}/2\pi = 3.33$\,kHz and vary the time $t_\text{del}$ between trigger synchronization and when the microwave sequence is applied, effectively changing the phase of the ac field sampled in the DRUM measurement. (b) Measured DRUM fringes, with sinusoidal fits, as a function of $B_y$ coil current for different delay times $t_\text{del}$. (c) Fitted $x$-coil and $y$-coil fringe frequencies for varied $t_\text{del}$. The almost $90^\circ$ phase shift between the traces is evidence of vector selectivity tunable with $t_\text{del}$. Error bars are standard error in fitted frequency.}
	\label{fig:figs3}
\end{figure}

We time the DRUM pulse sequence so that the $\pi$-pulse in the spin-echo sequence occurs at the same time as the zero-crossing of the AC field in the rotating frame. Practically, we determine this point by optimising the frequency of the DRUM fringes. Additionally, DRUM can be made sensitive to only $x$ or $y$ fields by adjusting the time between the motor synchronization signal and application of the microwave pulses. We confirm this behaviour by measuring DRUM fringes as a function of delay time $t_\text{del}$ and determining the fringe frequency. Fig \ref{fig:figs3} shows DRUM fringes at a measurement time of $\tau= 70\,\upmu$s for $x$-applied fields and $y$-applied fields as a function of $t_\text{del}$. The differing peak frequencies, expressed in mA$^{-1}$, reflect the different coil calibration constants. A degree of non-orthogonality is evident in our coils, as the $x$ and $y$ fringe frequency variation is $97^\circ$ out of phase, rather than exactly $90^\circ$. 

From these results, the known rotation speed and the known spin-echo time $\tau$, we can deduce at what delay time the zero-crossing occurs (in either rotation direction) and time sequences accordingly to maximize the DRUM sensitivity. While \emph{exclusive} $x$ or $y$-field sensitivity cannot be demonstrated with the coil geometry used in this work, for applied transverse fields exactly orthogonal to each other DRUM exhibits true vector sensitivity, adjusting the delay time can then be used to deduce the bearing of an applied magnetic test field.

\section{Calibration to absolute field}
When DRUM is synchronized to the rotation of the diamond, increasing $B_x$ or $B_y$ traces out monotonic sinusoidal fringes essentially indefinitely. When synchronization is absent, there is no well-defined relationship between the unconverted field and the measurement sequence. As such, the fringes phase-average to zero, except when the transverse fields are canceled to zero via application of unique $B_{x0}$ and $B_{y0}$ `nulling fields'. In this case, spin-echo contrast is maximized and sharply reduces as soon as the applied fields deviate from the nulling fields. This allows us to determine the currents corresponding to $B_{x0}$ and $B_{y0}$, and hence, calibrate DRUM fringes to an absolute field strength, namely, $B_x = B_y = 0$. 

\begin{figure}
	\centering
		\includegraphics[width = \columnwidth]{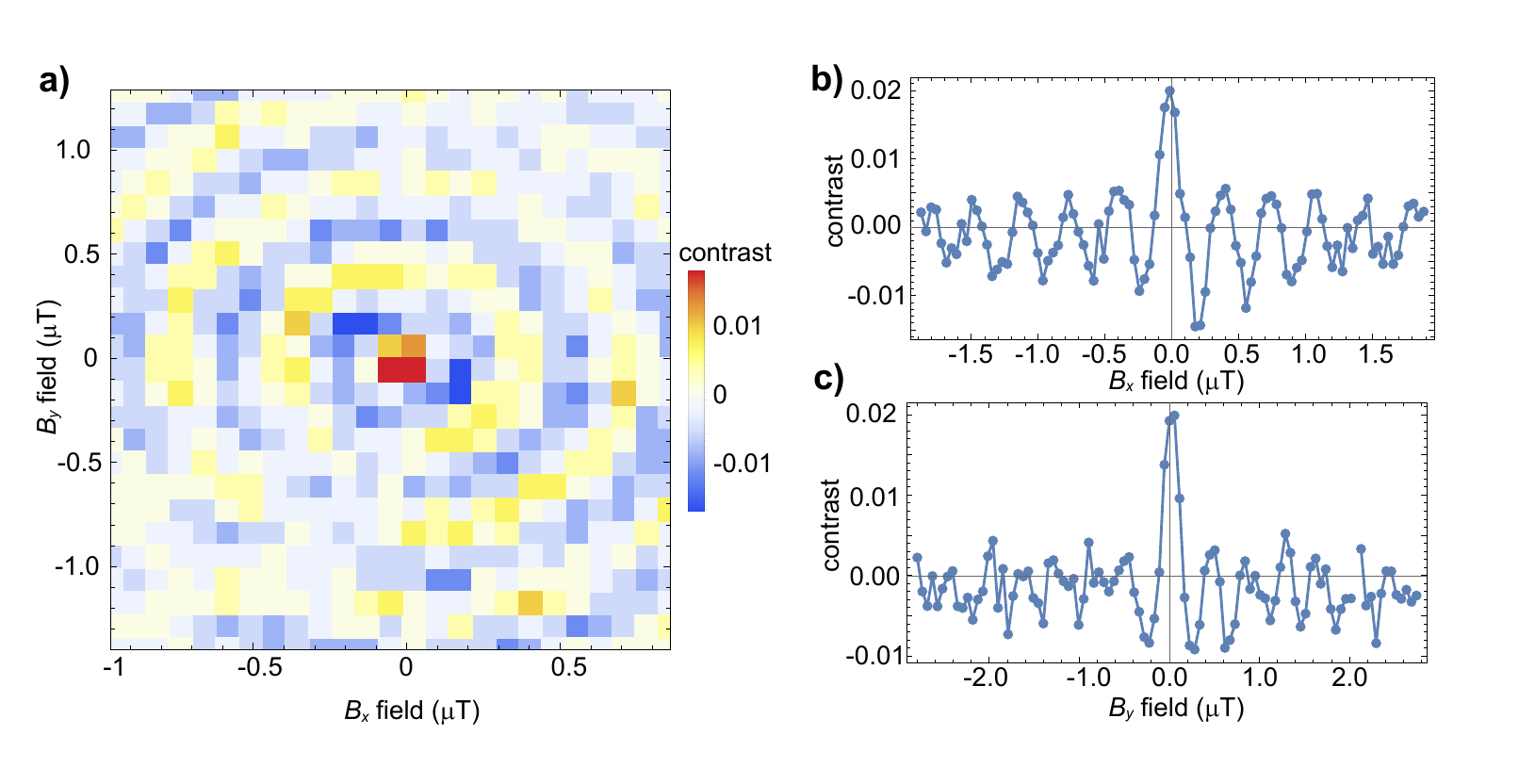}
	\caption{Calibration of DRUM to an absolute field. (a) when running asynchronously with the rotation ($\omega_\text{rot}/2\pi = 3.750\,$kHz, $\tau = 180\,\upmu$s), DRUM fringes rapidly damp, except when the transverse field is zero. Here, we vary the applied $B_x$ and $B_y$ fields, and find a sharp local maximum where the DRUM contrast peaks, which corresponds to $B_{x0} = B_{y0} = 0$, \emph{i.e.} the magnetic field is parallel to the rotation axis. (b) $B_x$ and (c) $B_y$ detailed optimisation, showing optimum contrast point and damping DRUM fringes.}
	\label{fig:figs4}
\end{figure}

Figure \ref{fig:figs4} shows the results of this calibration, for $\omega_\text{rot}/2\pi = 3.750\,$kHz and $\tau = 180\,\upmu$s, clearly indicating the values of $B_{x0}$ and $B_{y0}$. Much of the off-axis fields responsible for needing $B_{x0}$ and $B_{y0}$ in the first place is due to a slight tilt in the $z$-coil orientation, making it not perfectly parallel to the rotation axis. The required $B_{x0}$ and $B_{y0}$ therefore change with applied $z$-current, and we adjust $B_{x0}$ and $B_{y0}$ accordingly when changing $\tau$ (which requires variation of $B_z$), as discussed in the main text. The test fields applied with the $x$-coils in our demonstration of DRUM are therefore centered approximately at $B_x = 0$. In practice, we observe synchronized DRUM signals phase shifted from zero at $B_x = B_{x0}$, possibly due to drifting nulling field calibrations or imprecise knowledge of $B_{x0}$ and $B_{y0}$ in the first place: the fringe period when synchronised is much faster than the observed ripples in Fig. \ref{fig:figs4}, so small errors in $B_{x0}$ and $B_{y0}$ translate to larger phase offsets when synchronized.  

\section{Sensitivity derivation}
We follow the example of Ref. \cite{degen_quantum_2017s} in our derivation of the shot-noise limited DRUM sensitivity. We begin with the rotating NV, represented by the unit vector
\begin{equation}
\mathbf{n}(t) = \left(\sin\theta_\text{NV}\cos\left(\omega_\text{rot}t - \phi_0\right), \sin\theta_\text{NV}\sin\left(\omega_\text{rot}t - \phi_0\right), \cos\theta_\text{NV}\right),
\end{equation}
with $\theta_\text{NV}$ the polar angle between the NV and rotation axis $z$, $\phi_0$ set by the initial orientation of the NV axis and $\omega_\text{rot}$ the rotation angular frequency. We consider a static magnetic field in the stationary (lab) frame, 
\begin{equation}
\mathbf{B} = \left(B_x, 0, B_z\right)
\end{equation}
where, without loss of generality we have assumed $B_\perp = B_x$. The Zeeman shift experienced by the NV is well approximated for weak $B_x$ and $\theta_\text{NV}\ll 90^\circ$ by ($\hbar = 1$)
\begin{eqnarray}
E(m_S)&= & m_S \gamma_e \mathbf{B}\cdot\mathbf{n}(t)\nonumber\\
			&= & m_S\gamma_e B_z \cos\theta_\text{NV} + m_S \gamma_e B_x\sin\theta_\text{NV}\cos\left(\omega_\text{rot} t-\phi_0\right),
\label{eq:ssd}
\end{eqnarray}
with $m_S$ the magnetic quantum number of the NV and $\gamma_e/2\pi = 28\,$kHz/$\upmu$T. We consider microwaves resonant only with the $m_S = 0\leftrightarrow m_S = -1$ transition, in which case the time-dependent part of the transition energy is the second term in Eq. \ref{eq:ssd} with $m_S = -1$. In a spin-echo experiment with inter-pulse time $\tau$, the NV eigenstates will accumulate a relative phase 
\begin{equation}
\Phi = \int_{-\tau/2}^{0}E(t) dt - \int_{0}^{\tau/2}E(t) dt.
\label{eq:int0}
\end{equation}
Setting $\phi_0 = \pi/2$ ensures symmetric integration for $\tau$ either side of the zero crossing of the modulated energy, and this yields the optimum phase accumulation as a function of applied magnetic field. Equation \ref{eq:int0} evaluates to 
\begin{equation}
\Phi = \frac{4B_x\gamma_e}{\omega_\text{rot}}\sin\theta_\text{NV}\sin^2\left(\frac{\omega_\text{rot}\tau}{4}\right).
\label{eq:int1}
\end{equation}
Readout with a single $\pi/2$-pulse yields an NV fluorescence signal that varies as $\mathcal{S}_{\pi/2} = \cos^2\left(\Phi/2\right)$, while computing the (normalised) difference using a second sequence with a $180^\circ$-phase shifted (\emph{i.e.} -x) $\pi/2$-pulse yields $\mathcal{S} =\mathcal{S}_{\pi/2} - \mathcal{S}_{-\pi/2}  = \cos^2\left(\Phi/2\right) - \sin^2\left(\Phi/2\right) = \cos\Phi$. Operating at the mid-fringe point of the DRUM signal amounts to adding a given $B_x$ to induce a $\pi/2$ phase shift, \emph{i.e.} so that $\mathcal{S} = \cos(\Phi+\pi/2) = \sin\Phi\approx \Phi$. The transduction parameter, or slope $d\mathcal{S}/dB_x$, which converts magnetic field to quantum sensor readout is then
\begin{equation}
\frac{d\mathcal{S}}{dB_x} = \frac{4\gamma_e}{\omega_\text{rot}}\sin\theta_\text{NV}\sin^2\left(\frac{\omega_\text{rot}\tau}{4}\right).
\label{eq:int2}
\end{equation}

The sensitivity per unit bandwidth is the smallest detectable field $\delta B_x = \delta B$ that yields unity signal-to-noise in one second of averaging time. Regardless of $\tau$, we can perform a maximum of $1/t_\text{rot}$ measurements in one second. Thus
\begin{equation}
SNR = \delta B e^{-\left(\frac{\tau}{T_2}\right)^n}\frac{d\mathcal{S}}{dB} \frac{2C}{\sqrt{t_\text{rot}}}, 
\end{equation}
with $T_2$ the spin coherence time and $C = (1+4/(\epsilon^2 \mathcal{N} t_L))^{-1}\approx 0.1$ the readout efficiency; $t_L = 500\,$ns the laser pulse integration region, $\mathcal{N} = 9\times10^6$\,photons/s the steady-state count rate and $\epsilon = 0.1$ the maximum readout contrast ($\tau = 0$).  

Setting $SNR = 1$, substituting $\omega_\text{rot} = 2\pi/t_\text{rot}$ and rearranging for $\delta B$,
\begin{equation}
\delta B = \frac{\pi\,e^{\left(\frac{\tau}{T_2}\right)^n}}{4C\gamma_e\sin\theta_\text{NV}\sin^2\left(\frac{\pi \tau}{2 t_\text{rot}}\right)\sqrt{t_\text{rot}}}.
\end{equation}
The sensitivity is reduced by a factor of $2$, on account of self-normalisation to the tail of each PL measurement ($\sqrt{2}$) and halved duty cycle for normalisation with the $-\pi/2$-pulse ($\sqrt{2}$). However, due to $n = 3$ in the decay exponent, and as we have found from experiment, $\tau = T_2$ is not the optimum measurement time where the peak slope occurs at. Instead, we find typically $\tau_\text{opt} \sim 0.7-0.8\,T_2$, meaning the sensitivity reduces by a factor $e^{(0.7)^3}\approx e/2$ (at 3.750\,kHz). The shot-noise limited sensitivity is then

\begin{equation}
\delta B = \frac{\pi e}{4C\gamma_e\sin\theta_\text{NV}\sin^2\left(\frac{\pi \tau}{2 t_\text{rot}}\right)\sqrt{t_\text{rot}}}.
\label{eq:drumsens}
\end{equation}
For our current experimental parameters at $3.750$\,kHz, \emph{i.e.} $\theta_\text{NV} = 30^\circ$, $\tau = 180\,\upmu$s, $t_\text{rot}= 267\,\upmu$s, this evaluates to $18\,$nT\,Hz$^{-1}$.

The operating Ramsey sensitivity, which includes the same scaling due to self-normalisation and normalisation with the $-\pi/2$-pulse (multiplication by 2) is given by 
\begin{equation}
\delta B_\text{Ramsey} = \frac{\exp\left(\tau/T_2^\ast\right)}{\gamma_e C}\frac{\sqrt{\tau + t_D}}{\tau},
\label{eq:rs}
\end{equation} 
with $C$ the same as for DRUM, $T_2^\ast = 360\,$ns, $\tau$ the Ramsey time and $t_D$ the dead time, which includes a $3\,\upmu$s laser pulse, $1.2\,\upmu$s of dark time to maximise optical pumping efficiency due to spin state shelving and the combined durations of the two microwave pulses, $2t_{\pi/2} = t_\pi = 200\,$ns. The idealised, zero-dead time Ramsey sensitivity with which we quote comparison with in the main text is given by Eq. \ref{eq:rs} with $\tau = T_2^\ast$, $t_D = 0$ and is given by
\begin{equation}
\delta B_\text{Ramsey, ideal} = \frac{e}{2\gamma_e C}\frac{1}{\sqrt{T_2^\ast}}.
\label{eq:rs2}
\end{equation} 

The gain of DRUM over Ramsey is computed by taking the ratio of $\delta B_\text{DRUM}/\delta B_\text{Ramsey}$, with $\theta = 90^\circ$, $\tau = T_2 = t_\text{rot}$ in Eq. \ref{eq:drumsens} and the idealised shot-noise limited Ramsey sensitivity in Eq. \ref{eq:rs2}. We assume $C$ to be the same for both measurements. 

\section{The effects of rotation direction}
\label{sec:rot}
We also examined the effect of changing the direction of the motor's rotation. The data presented in the main text was using a rotation direction that created an effective magnetic `pseudo-field'~\cite{wood_magnetic_2017s} $B_\omega = \frac{\omega_\text{rot}}{\gamma_{13}} > 0$ ($\gamma_{13} = 10.71\,\text{Hz\,}\upmu\text{T}^{-1}$) that added to the $z$-oriented magnetic bias field, \emph{i.e.}
\begin{equation}
\mathbf{B} = B\,\uvect{z} = \left(B_z + B_\omega\right)\uvect{z}.
\label{eq:dd}
\end{equation} 

Importantly, this has the consequence of shifting $^{13}$C contrast revivals to \emph{shorter} times for an applied magnetic field. Therefore, to achieve a particular spin-echo measurement time $\tau$ at the first carbon-13 revival, a lower magnetic field is required
\begin{equation}
\tau= \frac{4\pi}{\gamma_{13}(B_z + B_\omega)}.
\label{eq:ddd}
\end{equation}

\begin{figure}
	\centering
		\includegraphics[width = \columnwidth]{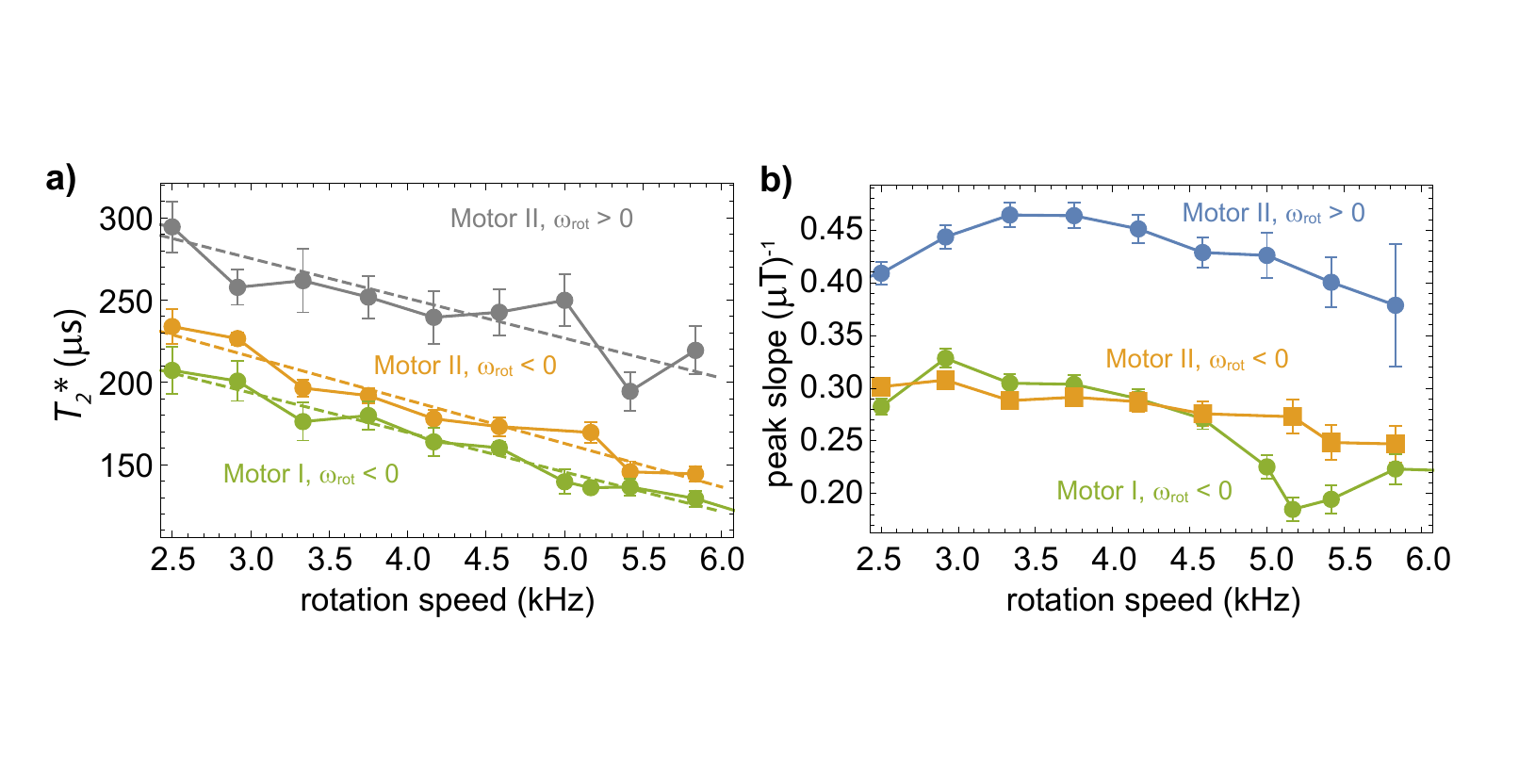}
	\caption{The effects of changing rotation direction, and instability in motor rotation. (a) $T_2$ as a function of rotation speed for two motors, Motor I and Motor II, each with a different diamond sample mounted. We observe a markedly higher $T_2$ for rotation against the direction of Larmor precession. (b) Peak sensing slopes for the same motor and diamond configurations in (a). A sharp dip near 5\,kHz in Motor I we believe is evidence for unstable rotation.}
	\label{fig:figs5}
\end{figure}

We found that when the rotation direction was reversed, and larger $z$-fields required for the same revival time, $T_2$ was lower. Figure \ref{fig:figs5} shows the pertinent data, as well as the slope $d\mathcal{S}/dB$ as a function of rotation speed. Different diamond samples were attached to Motor I and Motor II, though both samples exhibited nearly identical count rates, NV orientations and coherence times. In all cases, we observe $T_2$ to fall with rotation speed, at a similar rate, though the starting values are quite different, and these differences are reflected in the peak sensing slope data. It was found also that Motor I exhibited some form of instability near $5\,$kHz, leading to a marked drop in sensing slope (this observation led to the implementation of Motor II). We suspect motor run-in at a fixed speed (in this case, 3.333\,kHz) as a potential culprit for this instability.

Without $\omega_\text{rot} > 0$ data for Motor I allowing a conclusive determination, we posit with some confidence that the different coherence times for different rotation directions are due to the magnitude of $B_z$. While rotationally-symmetric, the $z$-bias field still makes an angle of $30^\circ$ to the NV axis. Larger $B$ fields tilted away from the NV axis induce more electron-spin state mixing, and, due to the anisotropy of the NV-$^{13}$C nuclear spin hyperfine interaction, a spread in nuclear spin precession frequencies~\cite{stanwix_coherence_2010s}. Consequently, the amplitude of the spin-echo signal is suppressed, and $T_2$ reduced. In other experiments (data not shown), we kept the spin-echo time $\tau$ constant but changed the magnetic field strength and order of the $^{13}$C revival we perform the DRUM measurement at, \emph{i.e.} for $\omega_\text{rot}/2\pi = 3.750\,$kHz, $\tau = 80\,\upmu$s and $B = 2.68\,$mT ($n = 1$) to $B = 9.0\,$mT ($n = 4$), and observed that higher fields indeed reduced the revival amplitude.  

\begin{thebibliography}{45}%
\makeatletter
\providecommand \@ifxundefined [1]{%
 \@ifx{#1\undefined}
}%
\providecommand \@ifnum [1]{%
 \ifnum #1\expandafter \@firstoftwo
 \else \expandafter \@secondoftwo
 \fi
}%
\providecommand \@ifx [1]{%
 \ifx #1\expandafter \@firstoftwo
 \else \expandafter \@secondoftwo
 \fi
}%
\providecommand \natexlab [1]{#1}%
\providecommand \enquote  [1]{``#1''}%
\providecommand \bibnamefont  [1]{#1}%
\providecommand \bibfnamefont [1]{#1}%
\providecommand \citenamefont [1]{#1}%
\providecommand \href@noop [0]{\@secondoftwo}%
\providecommand \href [0]{\begingroup \@sanitize@url \@href}%
\providecommand \@href[1]{\@@startlink{#1}\@@href}%
\providecommand \@@href[1]{\endgroup#1\@@endlink}%
\providecommand \@sanitize@url [0]{\catcode `\\12\catcode `\$12\catcode
  `\&12\catcode `\#12\catcode `\^12\catcode `\_12\catcode `\%12\relax}%
\providecommand \@@startlink[1]{}%
\providecommand \@@endlink[0]{}%
\providecommand \url  [0]{\begingroup\@sanitize@url \@url }%
\providecommand \@url [1]{\endgroup\@href {#1}{\urlprefix }}%
\providecommand \urlprefix  [0]{URL }%
\providecommand \Eprint [0]{\href }%
\providecommand \doibase [0]{https://doi.org/}%
\providecommand \selectlanguage [0]{\@gobble}%
\providecommand \bibinfo  [0]{\@secondoftwo}%
\providecommand \bibfield  [0]{\@secondoftwo}%
\providecommand \translation [1]{[#1]}%
\providecommand \BibitemOpen [0]{}%
\providecommand \bibitemStop [0]{}%
\providecommand \bibitemNoStop [0]{.\EOS\space}%
\providecommand \EOS [0]{\spacefactor3000\relax}%
\providecommand \BibitemShut  [1]{\csname bibitem#1\endcsname}%
\let\auto@bib@innerbib\@empty
\bibitem [{\citenamefont {Doherty}\ \emph {et~al.}(2013)\citenamefont
  {Doherty}, \citenamefont {Manson}, \citenamefont {Delaney}, \citenamefont
  {Jelezko}, \citenamefont {Wrachtrup},\ and\ \citenamefont
  {Hollenberg}}]{doherty_nitrogen-vacancy_2013}%
  \BibitemOpen
  \bibfield  {author} {\bibinfo {author} {\bibfnamefont {M.~W.}\ \bibnamefont
  {Doherty}}, \bibinfo {author} {\bibfnamefont {N.~B.}\ \bibnamefont {Manson}},
  \bibinfo {author} {\bibfnamefont {P.}~\bibnamefont {Delaney}}, \bibinfo
  {author} {\bibfnamefont {F.}~\bibnamefont {Jelezko}}, \bibinfo {author}
  {\bibfnamefont {J.}~\bibnamefont {Wrachtrup}},\ and\ \bibinfo {author}
  {\bibfnamefont {L.~C.~L.}\ \bibnamefont {Hollenberg}},\ }\bibfield  {title}
  \href
  {https://doi.org/10.1016/j.physrep.2013.02.001} {\bibfield  {journal}
  {\bibinfo  {journal} {Physics Reports}\ }\ \textbf {\bibinfo {volume}
  {528}},\ \bibinfo {pages} {1} (\bibinfo {year} {2013})}\BibitemShut {NoStop}%
\bibitem [{\citenamefont {Schirhagl}\ \emph {et~al.}(2014)\citenamefont
  {Schirhagl}, \citenamefont {Chang}, \citenamefont {Loretz},\ and\
  \citenamefont {Degen}}]{schirhagl_nitrogen-vacancy_2014}%
  \BibitemOpen
  \bibfield  {author} {\bibinfo {author} {\bibfnamefont {R.}~\bibnamefont
  {Schirhagl}}, \bibinfo {author} {\bibfnamefont {K.}~\bibnamefont {Chang}},
  \bibinfo {author} {\bibfnamefont {M.}~\bibnamefont {Loretz}},\ and\ \bibinfo
  {author} {\bibfnamefont {C.~L.}\ \bibnamefont {Degen}},\ }
  \href{https://doi.org/10.1146/annurev-physchem-040513-103659} {\bibfield
  {journal} {\bibinfo  {journal} {Annual Review of Physical Chemistry}\
  }\textbf {\bibinfo {volume} {65}},\ \bibinfo {pages} {83} (\bibinfo {year}
  {2014})}\BibitemShut {NoStop}%
\bibitem [{\citenamefont {Gille}\ \emph {et~al.}(2021)\citenamefont {Gille},
  \citenamefont {McCoey}, \citenamefont {Hall}, \citenamefont {Tetienne},
  \citenamefont {Malkemper}, \citenamefont {Keays}, \citenamefont
  {Hollenberg},\ and\ \citenamefont {Simpson}}]{gille_quantum_2021}%
  \BibitemOpen
  \bibfield  {author} {\bibinfo {author} {\bibfnamefont {R.~W.~d.}\
  \bibnamefont {Gille}}, \bibinfo {author} {\bibfnamefont {J.~M.}\ \bibnamefont
  {McCoey}}, \bibinfo {author} {\bibfnamefont {L.~T.}\ \bibnamefont {Hall}},
  \bibinfo {author} {\bibfnamefont {J.-P.}\ \bibnamefont {Tetienne}}, \bibinfo
  {author} {\bibfnamefont {E.~P.}\ \bibnamefont {Malkemper}}, \bibinfo {author}
  {\bibfnamefont {D.~A.}\ \bibnamefont {Keays}}, \bibinfo {author}
  {\bibfnamefont {L.~C.~L.}\ \bibnamefont {Hollenberg}},\ and\ \bibinfo
  {author} {\bibfnamefont {D.~A.}\ \bibnamefont {Simpson}},\ }\href
  {https://doi.org/10.1073/pnas.2112749118} {PNAS \textbf {118} e2112749118 (2021)}\BibitemShut {NoStop}%
\bibitem [{\citenamefont {Tetienne}\ \emph {et~al.}(2017)\citenamefont
  {Tetienne}, \citenamefont {Dontschuk}, \citenamefont {Broadway},
  \citenamefont {Stacey}, \citenamefont {Simpson},\ and\ \citenamefont
  {Hollenberg}}]{tetienne_quantum_2017}%
  \BibitemOpen
  \bibfield  {author} {\bibinfo {author} {\bibfnamefont {J.-P.}\ \bibnamefont
  {Tetienne}}, \bibinfo {author} {\bibfnamefont {N.}~\bibnamefont {Dontschuk}},
  \bibinfo {author} {\bibfnamefont {D.~A.}\ \bibnamefont {Broadway}}, \bibinfo
  {author} {\bibfnamefont {A.}~\bibnamefont {Stacey}}, \bibinfo {author}
  {\bibfnamefont {D.~A.}\ \bibnamefont {Simpson}},\ and\ \bibinfo {author}
  {\bibfnamefont {L.~C.~L.}\ \bibnamefont {Hollenberg}},\ }
  \href {https://doi.org/10.1126/sciadv.1602429} {\bibfield
  {journal} {\bibinfo  {journal} {Science Advances}\ }\textbf {\bibinfo
  {volume} {3}},\ \bibinfo {pages} {e1602429} (\bibinfo {year} {2017})}\BibitemShut {NoStop}%
\bibitem [{\citenamefont {Casola}\ \emph {et~al.}(2018)\citenamefont {Casola},
  \citenamefont {van~der Sar},\ and\ \citenamefont
  {Yacoby}}]{casola_probing_2018}%
  \BibitemOpen
  \bibfield  {author} {\bibinfo {author} {\bibfnamefont {F.}~\bibnamefont
  {Casola}}, \bibinfo {author} {\bibfnamefont {T.}~\bibnamefont {van~der
  Sar}},\ and\ \bibinfo {author} {\bibfnamefont {A.}~\bibnamefont {Yacoby}},\
  }\href {https://doi.org/10.1038/natrevmats.2017.88} {\bibfield
  {journal} {\bibinfo  {journal} {Nat Rev Mater}\ }\textbf {\bibinfo {volume}
  {3}},\ \bibinfo {pages} {1} (\bibinfo {year} {2018})}\BibitemShut {NoStop}%
\bibitem [{\citenamefont {Mamin}\ \emph {et~al.}(2013)\citenamefont {Mamin},
  \citenamefont {Kim}, \citenamefont {Sherwood}, \citenamefont {Rettner},
  \citenamefont {Ohno}, \citenamefont {Awschalom},\ and\ \citenamefont
  {Rugar}}]{mamin_nanoscale_2013}%
  \BibitemOpen
  \bibfield  {author} {\bibinfo {author} {\bibfnamefont {H.~J.}\ \bibnamefont
  {Mamin}}, \bibinfo {author} {\bibfnamefont {M.}~\bibnamefont {Kim}}, \bibinfo
  {author} {\bibfnamefont {M.~H.}\ \bibnamefont {Sherwood}}, \bibinfo {author}
  {\bibfnamefont {C.~T.}\ \bibnamefont {Rettner}}, \bibinfo {author}
  {\bibfnamefont {K.}~\bibnamefont {Ohno}}, \bibinfo {author} {\bibfnamefont
  {D.~D.}\ \bibnamefont {Awschalom}},\ and\ \bibinfo {author} {\bibfnamefont
  {D.}~\bibnamefont {Rugar}},\ }\href{https://doi.org/10.1126/science.1231540} {\bibfield  {journal} {\bibinfo
  {journal} {Science}\ }\textbf {\bibinfo {volume} {339}},\ \bibinfo {pages}
  {557} (\bibinfo {year} {2013})}\BibitemShut
  {NoStop}%
\bibitem [{\citenamefont {Fu}\ \emph {et~al.}(2020)\citenamefont {Fu},
  \citenamefont {Iwata}, \citenamefont {Wickenbrock},\ and\ \citenamefont
  {Budker}}]{fu_sensitive_2020}%
  \BibitemOpen
  \bibfield  {author} {\bibinfo {author} {\bibfnamefont {K.-M.~C.}\
  \bibnamefont {Fu}}, \bibinfo {author} {\bibfnamefont {G.~Z.}\ \bibnamefont
  {Iwata}}, \bibinfo {author} {\bibfnamefont {A.}~\bibnamefont {Wickenbrock}},\
  and\ \bibinfo {author} {\bibfnamefont {D.}~\bibnamefont {Budker}},\
  }\href {https://doi.org/10.1116/5.0025186} {\bibfield
  {journal} {\bibinfo  {journal} {AVS Quantum Sci.}\ }\textbf {\bibinfo
  {volume} {2}},\ \bibinfo {pages} {044702} (\bibinfo {year} {2020})}\BibitemShut {NoStop}%
\bibitem [{\citenamefont {Barry}\ \emph {et~al.}(2020)\citenamefont {Barry},
  \citenamefont {Schloss}, \citenamefont {Bauch}, \citenamefont {Turner},
  \citenamefont {Hart}, \citenamefont {Pham},\ and\ \citenamefont
  {Walsworth}}]{barry_sensitivity_2020}%
  \BibitemOpen
  \bibfield  {author} {\bibinfo {author} {\bibfnamefont {J.~F.}\ \bibnamefont
  {Barry}}, \bibinfo {author} {\bibfnamefont {J.~M.}\ \bibnamefont {Schloss}},
  \bibinfo {author} {\bibfnamefont {E.}~\bibnamefont {Bauch}}, \bibinfo
  {author} {\bibfnamefont {M.~J.}\ \bibnamefont {Turner}}, \bibinfo {author}
  {\bibfnamefont {C.~A.}\ \bibnamefont {Hart}}, \bibinfo {author}
  {\bibfnamefont {L.~M.}\ \bibnamefont {Pham}},\ and\ \bibinfo {author}
  {\bibfnamefont {R.~L.}\ \bibnamefont {Walsworth}},\ }\href {https://doi.org/10.1103/RevModPhys.92.015004} {\bibfield  {journal}
  {\bibinfo  {journal} {Rev. Mod. Phys.}\ }\textbf {\bibinfo {volume} {92}},\
  \bibinfo {pages} {015004} (\bibinfo {year} {2020})}\BibitemShut {NoStop}%
\bibitem [{\citenamefont {Balasubramanian}\ \emph {et~al.}(2009)\citenamefont
  {Balasubramanian}, \citenamefont {Neumann}, \citenamefont {Twitchen},
  \citenamefont {Markham}, \citenamefont {Kolesov}, \citenamefont {Mizuochi},
  \citenamefont {Isoya}, \citenamefont {Achard}, \citenamefont {Beck},
  \citenamefont {Tissler}, \citenamefont {Jacques}, \citenamefont {Hemmer},
  \citenamefont {Jelezko},\ and\ \citenamefont
  {Wrachtrup}}]{balasubramanian_ultralong_2009}%
  \BibitemOpen
  \bibfield  {author} {\bibinfo {author} {\bibfnamefont {G.}~\bibnamefont
  {Balasubramanian}}, \bibinfo {author} {\bibfnamefont {P.}~\bibnamefont
  {Neumann}}, \bibinfo {author} {\bibfnamefont {D.}~\bibnamefont {Twitchen}},
  \bibinfo {author} {\bibfnamefont {M.}~\bibnamefont {Markham}}, \bibinfo
  {author} {\bibfnamefont {R.}~\bibnamefont {Kolesov}}, \bibinfo {author}
  {\bibfnamefont {N.}~\bibnamefont {Mizuochi}}, \bibinfo {author}
  {\bibfnamefont {J.}~\bibnamefont {Isoya}}, \bibinfo {author} {\bibfnamefont
  {J.}~\bibnamefont {Achard}}, \bibinfo {author} {\bibfnamefont
  {J.}~\bibnamefont {Beck}}, \bibinfo {author} {\bibfnamefont {J.}~\bibnamefont
  {Tissler}}, \bibinfo {author} {\bibfnamefont {V.}~\bibnamefont {Jacques}},
  \bibinfo {author} {\bibfnamefont {P.~R.}\ \bibnamefont {Hemmer}}, \bibinfo
  {author} {\bibfnamefont {F.}~\bibnamefont {Jelezko}},\ and\ \bibinfo {author}
  {\bibfnamefont {J.}~\bibnamefont {Wrachtrup}},\ }\href {https://doi.org/10.1038/nmat2420}
  {\bibfield  {journal} {\bibinfo  {journal} {Nat Mater}\ }\textbf {\bibinfo
  {volume} {8}},\ \bibinfo {pages} {383} (\bibinfo {year} {2009})}\BibitemShut {NoStop}%
\bibitem [{\citenamefont {Herbschleb}\ \emph {et~al.}(2019)\citenamefont
  {Herbschleb}, \citenamefont {Kato}, \citenamefont {Maruyama}, \citenamefont
  {Danjo}, \citenamefont {Makino}, \citenamefont {Yamasaki}, \citenamefont
  {Ohki}, \citenamefont {Hayashi}, \citenamefont {Morishita}, \citenamefont
  {Fujiwara},\ and\ \citenamefont {Mizuochi}}]{herbschleb_ultra-long_2019}%
  \BibitemOpen
  \bibfield  {author} {\bibinfo {author} {\bibfnamefont {E.~D.}\ \bibnamefont
  {Herbschleb}}, \bibinfo {author} {\bibfnamefont {H.}~\bibnamefont {Kato}},
  \bibinfo {author} {\bibfnamefont {Y.}~\bibnamefont {Maruyama}}, \bibinfo
  {author} {\bibfnamefont {T.}~\bibnamefont {Danjo}}, \bibinfo {author}
  {\bibfnamefont {T.}~\bibnamefont {Makino}}, \bibinfo {author} {\bibfnamefont
  {S.}~\bibnamefont {Yamasaki}}, \bibinfo {author} {\bibfnamefont
  {I.}~\bibnamefont {Ohki}}, \bibinfo {author} {\bibfnamefont {K.}~\bibnamefont
  {Hayashi}}, \bibinfo {author} {\bibfnamefont {H.}~\bibnamefont {Morishita}},
  \bibinfo {author} {\bibfnamefont {M.}~\bibnamefont {Fujiwara}},\ and\
  \bibinfo {author} {\bibfnamefont {N.}~\bibnamefont {Mizuochi}},\ }\href
  {https://doi.org/10.1038/s41467-019-11776-8} {\bibfield  {journal} {\bibinfo
  {journal} {Nat Commun}\ }\textbf {\bibinfo {volume} {10}},\ \bibinfo {pages}
  {3766} (\bibinfo {year} {2019})}\BibitemShut{NoStop}%
\bibitem [{\citenamefont {Clevenson}\ \emph {et~al.}(2015)\citenamefont
  {Clevenson}, \citenamefont {Trusheim}, \citenamefont {Teale}, \citenamefont
  {Schröder}, \citenamefont {Braje},\ and\ \citenamefont
  {Englund}}]{clevenson_broadband_2015}%
  \BibitemOpen
  \bibfield  {author} {\bibinfo {author} {\bibfnamefont {H.}~\bibnamefont
  {Clevenson}}, \bibinfo {author} {\bibfnamefont {M.~E.}\ \bibnamefont
  {Trusheim}}, \bibinfo {author} {\bibfnamefont {C.}~\bibnamefont {Teale}},
  \bibinfo {author} {\bibfnamefont {T.}~\bibnamefont {Schröder}}, \bibinfo
  {author} {\bibfnamefont {D.}~\bibnamefont {Braje}},\ and\ \bibinfo {author}
  {\bibfnamefont {D.}~\bibnamefont {Englund}},\ }\href
  {https://doi.org/10.1038/nphys3291} {\bibfield  {journal} {\bibinfo
  {journal} {Nat. Phys.}\ }\textbf {\bibinfo {volume} {11}},\ \bibinfo
  {pages} {393} (\bibinfo {year} {2015})}\BibitemShut {NoStop}%
\bibitem [{\citenamefont {Lange}\ \emph {et~al.}(2012)\citenamefont {Lange},
  \citenamefont {Sar}, \citenamefont {Blok}, \citenamefont {Wang},
  \citenamefont {Dobrovitski},\ and\ \citenamefont
  {Hanson}}]{lange_controlling_2012}%
  \BibitemOpen
  \bibfield  {author} {\bibinfo {author} {\bibfnamefont {G.~d.}\ \bibnamefont
  {Lange}}, \bibinfo {author} {\bibfnamefont {T.~v.~d.}\ \bibnamefont {Sar}},
  \bibinfo {author} {\bibfnamefont {M.}~\bibnamefont {Blok}}, \bibinfo {author}
  {\bibfnamefont {Z.-H.}\ \bibnamefont {Wang}}, \bibinfo {author}
  {\bibfnamefont {V.}~\bibnamefont {Dobrovitski}},\ and\ \bibinfo {author}
  {\bibfnamefont {R.}~\bibnamefont {Hanson}},\ }\href
  {https://doi.org/10.1038/srep00382} {\bibfield  {journal} {\bibinfo
  {journal} {Sci. Rep.}\ }\textbf {\bibinfo {volume} {2}},\ \bibinfo
  {pages} {382} (\bibinfo {year} {2012})}\BibitemShut {NoStop}%
\bibitem [{\citenamefont {Mamin}\ \emph {et~al.}(2014)\citenamefont {Mamin},
  \citenamefont {Sherwood}, \citenamefont {Kim}, \citenamefont {Rettner},
  \citenamefont {Ohno}, \citenamefont {Awschalom},\ and\ \citenamefont
  {Rugar}}]{mamin_multipulse_2014}%
  \BibitemOpen
  \bibfield  {author} {\bibinfo {author} {\bibfnamefont {H.}~\bibnamefont
  {Mamin}}, \bibinfo {author} {\bibfnamefont {M.}~\bibnamefont {Sherwood}},
  \bibinfo {author} {\bibfnamefont {M.}~\bibnamefont {Kim}}, \bibinfo {author}
  {\bibfnamefont {C.}~\bibnamefont {Rettner}}, \bibinfo {author} {\bibfnamefont
  {K.}~\bibnamefont {Ohno}}, \bibinfo {author} {\bibfnamefont {D.}~\bibnamefont
  {Awschalom}},\ and\ \bibinfo {author} {\bibfnamefont {D.}~\bibnamefont
  {Rugar}},\ }\href {https://doi.org/10.1103/PhysRevLett.113.030803}
  {\bibfield  {journal} {\bibinfo  {journal} {Phys. Rev. Lett.}\ }\textbf
  {\bibinfo {volume} {113}},\ \bibinfo {pages} {030803} (\bibinfo {year}
  {2014})}\BibitemShut {NoStop}%
\bibitem [{\citenamefont {Bauch}\ \emph {et~al.}(2018)\citenamefont {Bauch},
  \citenamefont {Hart}, \citenamefont {Schloss}, \citenamefont {Turner},
  \citenamefont {Barry}, \citenamefont {Kehayias}, \citenamefont {Singh},\ and\
  \citenamefont {Walsworth}}]{bauch_ultralong_2018}%
  \BibitemOpen
  \bibfield  {author} {\bibinfo {author} {\bibfnamefont {E.}~\bibnamefont
  {Bauch}}, \bibinfo {author} {\bibfnamefont {C.~A.}\ \bibnamefont {Hart}},
  \bibinfo {author} {\bibfnamefont {J.~M.}\ \bibnamefont {Schloss}}, \bibinfo
  {author} {\bibfnamefont {M.~J.}\ \bibnamefont {Turner}}, \bibinfo {author}
  {\bibfnamefont {J.~F.}\ \bibnamefont {Barry}}, \bibinfo {author}
  {\bibfnamefont {P.}~\bibnamefont {Kehayias}}, \bibinfo {author}
  {\bibfnamefont {S.}~\bibnamefont {Singh}},\ and\ \bibinfo {author}
  {\bibfnamefont {R.~L.}\ \bibnamefont {Walsworth}},\ }\href
  {https://doi.org/10.1103/PhysRevX.8.031025} {\bibfield  {journal} {\bibinfo
  {journal} {Phys. Rev. X}\ }\textbf {\bibinfo {volume} {8}},\ \bibinfo {pages}
  {031025} (\bibinfo {year} {2018})}\BibitemShut
  {NoStop}%
\bibitem [{\citenamefont {Fescenko}\ \emph {et~al.}(2020)\citenamefont
  {Fescenko}, \citenamefont {Jarmola}, \citenamefont {Savukov}, \citenamefont
  {Kehayias}, \citenamefont {Smits}, \citenamefont {Damron}, \citenamefont
  {Ristoff}, \citenamefont {Mosavian},\ and\ \citenamefont
  {Acosta}}]{fescenko_diamond_2020}%
  \BibitemOpen
  \bibfield  {author} {\bibinfo {author} {\bibfnamefont {I.}~\bibnamefont
  {Fescenko}}, \bibinfo {author} {\bibfnamefont {A.}~\bibnamefont {Jarmola}},
  \bibinfo {author} {\bibfnamefont {I.}~\bibnamefont {Savukov}}, \bibinfo
  {author} {\bibfnamefont {P.}~\bibnamefont {Kehayias}}, \bibinfo {author}
  {\bibfnamefont {J.}~\bibnamefont {Smits}}, \bibinfo {author} {\bibfnamefont
  {J.}~\bibnamefont {Damron}}, \bibinfo {author} {\bibfnamefont
  {N.}~\bibnamefont {Ristoff}}, \bibinfo {author} {\bibfnamefont
  {N.}~\bibnamefont {Mosavian}},\ and\ \bibinfo {author} {\bibfnamefont
  {V.~M.}\ \bibnamefont {Acosta}},\ }\href
  {https://doi.org/10.1103/PhysRevResearch.2.023394} {\bibfield  {journal}
  {\bibinfo  {journal} {Phys. Rev. Research}\ }\textbf {\bibinfo {volume}
  {2}},\ \bibinfo {pages} {023394} (\bibinfo {year} {2020})}\BibitemShut {NoStop}%
\bibitem [{\citenamefont {Zhang}\ \emph {et~al.}(2021)\citenamefont {Zhang},
  \citenamefont {Shagieva}, \citenamefont {Widmann}, \citenamefont {Kübler},
  \citenamefont {Vorobyov}, \citenamefont {Kapitanova}, \citenamefont
  {Nenasheva}, \citenamefont {Corkill}, \citenamefont {Rhrle}, \citenamefont
  {Nakamura}, \citenamefont {Sumiya}, \citenamefont {Onoda}, \citenamefont
  {Isoya},\ and\ \citenamefont {Wrachtrup}}]{zhang_diamond_2021}%
  \BibitemOpen
  \bibfield  {author} {\bibinfo {author} {\bibfnamefont {C.}~\bibnamefont
  {Zhang}}, \bibinfo {author} {\bibfnamefont {F.}~\bibnamefont {Shagieva}},
  \bibinfo {author} {\bibfnamefont {M.}~\bibnamefont {Widmann}}, \bibinfo
  {author} {\bibfnamefont {M.}~\bibnamefont {Kübler}}, \bibinfo {author}
  {\bibfnamefont {V.}~\bibnamefont {Vorobyov}}, \bibinfo {author}
  {\bibfnamefont {P.}~\bibnamefont {Kapitanova}}, \bibinfo {author}
  {\bibfnamefont {E.}~\bibnamefont {Nenasheva}}, \bibinfo {author}
  {\bibfnamefont {R.}~\bibnamefont {Corkill}}, \bibinfo {author} {\bibfnamefont
  {O.}~\bibnamefont {Rhrle}}, \bibinfo {author} {\bibfnamefont
  {K.}~\bibnamefont {Nakamura}}, \bibinfo {author} {\bibfnamefont
  {H.}~\bibnamefont {Sumiya}}, \bibinfo {author} {\bibfnamefont
  {S.}~\bibnamefont {Onoda}}, \bibinfo {author} {\bibfnamefont
  {J.}~\bibnamefont {Isoya}},\ and\ \bibinfo {author} {\bibfnamefont
  {J.}~\bibnamefont {Wrachtrup}},\ }\href {https://doi.org/10.1103/PhysRevApplied.15.064075}
  {\bibfield  {journal} {\bibinfo  {journal} {Phys. Rev. Applied}\ }\textbf
  {\bibinfo {volume} {15}},\ \bibinfo {pages} {064075} (\bibinfo {year}
  {2021})}\BibitemShut
  {NoStop}%
\bibitem [{\citenamefont {Ramsey}(1950)}]{ramsey_molecular_1950}%
  \BibitemOpen
  \bibfield  {author} {\bibinfo {author} {\bibfnamefont {N.~F.}\ \bibnamefont
  {Ramsey}},\ }\href
  {https://doi.org/10.1103/PhysRev.78.695} {\bibfield  {journal} {\bibinfo
  {journal} {Phys. Rev.}\ }\textbf {\bibinfo {volume} {78}},\ \bibinfo {pages}
  {695} (\bibinfo {year} {1950})}\BibitemShut {NoStop}%
\bibitem [{\citenamefont {Rondin}\ \emph {et~al.}(2014)\citenamefont {Rondin},
  \citenamefont {Tetienne}, \citenamefont {Hingant}, \citenamefont {Roch},
  \citenamefont {Maletinsky},\ and\ \citenamefont
  {Jacques}}]{rondin_magnetometry_2014}%
  \BibitemOpen
  \bibfield  {author} {\bibinfo {author} {\bibfnamefont {L.}~\bibnamefont
  {Rondin}}, \bibinfo {author} {\bibfnamefont {J.-P.}\ \bibnamefont
  {Tetienne}}, \bibinfo {author} {\bibfnamefont {T.}~\bibnamefont {Hingant}},
  \bibinfo {author} {\bibfnamefont {J.-F.}\ \bibnamefont {Roch}}, \bibinfo
  {author} {\bibfnamefont {P.}~\bibnamefont {Maletinsky}},\ and\ \bibinfo
  {author} {\bibfnamefont {V.}~\bibnamefont {Jacques}},\ }\href {https://doi.org/10.1088/0034-4885/77/5/056503}
  {\bibfield  {journal} {\bibinfo  {journal} {Rep. Prog. Phys.}\ }\textbf
  {\bibinfo {volume} {77}},\ \bibinfo {pages} {056503} (\bibinfo {year}
  {2014})}\BibitemShut {NoStop}%
\bibitem [{\citenamefont {Childress}\ \emph {et~al.}(2006)\citenamefont
  {Childress}, \citenamefont {Dutt}, \citenamefont {Taylor}, \citenamefont
  {Zibrov}, \citenamefont {Jelezko}, \citenamefont {Wrachtrup}, \citenamefont
  {Hemmer},\ and\ \citenamefont {Lukin}}]{childress_coherent_2006}%
  \BibitemOpen
  \bibfield  {author} {\bibinfo {author} {\bibfnamefont {L.}~\bibnamefont
  {Childress}}, \bibinfo {author} {\bibfnamefont {M.~V.~G.}\ \bibnamefont
  {Dutt}}, \bibinfo {author} {\bibfnamefont {J.~M.}\ \bibnamefont {Taylor}},
  \bibinfo {author} {\bibfnamefont {A.~S.}\ \bibnamefont {Zibrov}}, \bibinfo
  {author} {\bibfnamefont {F.}~\bibnamefont {Jelezko}}, \bibinfo {author}
  {\bibfnamefont {J.}~\bibnamefont {Wrachtrup}}, \bibinfo {author}
  {\bibfnamefont {P.~R.}\ \bibnamefont {Hemmer}},\ and\ \bibinfo {author}
  {\bibfnamefont {M.~D.}\ \bibnamefont {Lukin}},\ }\href
  {https://doi.org/10.1126/science.1131871} {\bibfield  {journal} {\bibinfo
  {journal} {Science}\ }\textbf {\bibinfo {volume} {314}},\ \bibinfo {pages}
  {281} (\bibinfo {year} {2006})}\BibitemShut
  {NoStop}%
\bibitem [{\citenamefont {Bauch}\ \emph {et~al.}(2020)\citenamefont {Bauch},
  \citenamefont {Singh}, \citenamefont {Lee}, \citenamefont {Hart},
  \citenamefont {Schloss}, \citenamefont {Turner}, \citenamefont {Barry},
  \citenamefont {Pham}, \citenamefont {Bar-Gill}, \citenamefont {Yelin},\ and\
  \citenamefont {Walsworth}}]{bauch_decoherence_2020}%
  \BibitemOpen
  \bibfield  {author} {\bibinfo {author} {\bibfnamefont {E.}~\bibnamefont
  {Bauch}}, \bibinfo {author} {\bibfnamefont {S.}~\bibnamefont {Singh}},
  \bibinfo {author} {\bibfnamefont {J.}~\bibnamefont {Lee}}, \bibinfo {author}
  {\bibfnamefont {C.~A.}\ \bibnamefont {Hart}}, \bibinfo {author}
  {\bibfnamefont {J.~M.}\ \bibnamefont {Schloss}}, \bibinfo {author}
  {\bibfnamefont {M.~J.}\ \bibnamefont {Turner}}, \bibinfo {author}
  {\bibfnamefont {J.~F.}\ \bibnamefont {Barry}}, \bibinfo {author}
  {\bibfnamefont {L.~M.}\ \bibnamefont {Pham}}, \bibinfo {author}
  {\bibfnamefont {N.}~\bibnamefont {Bar-Gill}}, \bibinfo {author}
  {\bibfnamefont {S.~F.}\ \bibnamefont {Yelin}},\ and\ \bibinfo {author}
  {\bibfnamefont {R.~L.}\ \bibnamefont {Walsworth}},\ }\href {https://doi.org/10.1103/PhysRevB.102.134210} {\bibfield
  {journal} {\bibinfo  {journal} {Phys. Rev. B}\ }\textbf {\bibinfo {volume}
  {102}},\ \bibinfo {pages} {134210} (\bibinfo {year} {2020})}\BibitemShut {NoStop}%
\bibitem [{\citenamefont {Hahn}(1950)}]{hahn_spin_1950}%
  \BibitemOpen
  \bibfield  {author} {\bibinfo {author} {\bibfnamefont {E.~L.}\ \bibnamefont
  {Hahn}},\ }\href
  {https://doi.org/10.1103/PhysRev.80.580} {\bibfield  {journal} {\bibinfo
  {journal} {Phys. Rev.}\ }\textbf {\bibinfo {volume} {80}},\ \bibinfo {pages}
  {580} (\bibinfo {year} {1950})},\BibitemShut
  {NoStop}%
\bibitem [{\citenamefont {Taylor}\ \emph {et~al.}(2008)\citenamefont {Taylor},
  \citenamefont {Cappellaro}, \citenamefont {Childress}, \citenamefont {Jiang},
  \citenamefont {Budker}, \citenamefont {Hemmer}, \citenamefont {Yacoby},
  \citenamefont {Walsworth},\ and\ \citenamefont
  {Lukin}}]{taylor_high-sensitivity_2008}%
  \BibitemOpen
  \bibfield  {author} {\bibinfo {author} {\bibfnamefont {J.~M.}\ \bibnamefont
  {Taylor}}, \bibinfo {author} {\bibfnamefont {P.}~\bibnamefont {Cappellaro}},
  \bibinfo {author} {\bibfnamefont {L.}~\bibnamefont {Childress}}, \bibinfo
  {author} {\bibfnamefont {L.}~\bibnamefont {Jiang}}, \bibinfo {author}
  {\bibfnamefont {D.}~\bibnamefont {Budker}}, \bibinfo {author} {\bibfnamefont
  {P.~R.}\ \bibnamefont {Hemmer}}, \bibinfo {author} {\bibfnamefont
  {A.}~\bibnamefont {Yacoby}}, \bibinfo {author} {\bibfnamefont
  {R.}~\bibnamefont {Walsworth}},\ and\ \bibinfo {author} {\bibfnamefont
  {M.~D.}\ \bibnamefont {Lukin}},\ }\href {https://doi.org/10.1038/nphys1075} {\bibfield
  {journal} {\bibinfo  {journal} {Nat Phys}\ }\textbf {\bibinfo {volume} {4}},\
  \bibinfo {pages} {810} (\bibinfo {year} {2008})}\BibitemShut {NoStop}%
\bibitem [{\citenamefont {Acosta}\ \emph {et~al.}(2010)\citenamefont {Acosta},
  \citenamefont {Bauch}, \citenamefont {Jarmola}, \citenamefont {Zipp},
  \citenamefont {Ledbetter},\ and\ \citenamefont
  {Budker}}]{acosta_broadband_2010}%
  \BibitemOpen
  \bibfield  {author} {\bibinfo {author} {\bibfnamefont {V.~M.}\ \bibnamefont
  {Acosta}}, \bibinfo {author} {\bibfnamefont {E.}~\bibnamefont {Bauch}},
  \bibinfo {author} {\bibfnamefont {A.}~\bibnamefont {Jarmola}}, \bibinfo
  {author} {\bibfnamefont {L.~J.}\ \bibnamefont {Zipp}}, \bibinfo {author}
  {\bibfnamefont {M.~P.}\ \bibnamefont {Ledbetter}},\ and\ \bibinfo {author}
  {\bibfnamefont {D.}~\bibnamefont {Budker}},\ }\href
  {https://doi.org/10.1063/1.3507884} {\bibfield  {journal} {\bibinfo
  {journal} {Applied Physics Letters}\ }\textbf {\bibinfo {volume} {97}},\
  \bibinfo {pages} {174104} (\bibinfo {year} {2010})}\BibitemShut {NoStop}%
\bibitem [{\citenamefont {Jeske}\ \emph {et~al.}(2016)\citenamefont {Jeske},
  \citenamefont {Cole},\ and\ \citenamefont {Greentree}}]{jeske_laser_2016}%
  \BibitemOpen
  \bibfield  {author} {\bibinfo {author} {\bibfnamefont {J.}~\bibnamefont
  {Jeske}}, \bibinfo {author} {\bibfnamefont {J.~H.}\ \bibnamefont {Cole}},\
  and\ \bibinfo {author} {\bibfnamefont {A.~D.}\ \bibnamefont {Greentree}},\
  }\href {https://doi.org/10.1088/1367-2630/18/1/013015}
  {\bibfield  {journal} {\bibinfo  {journal} {New J. Phys.}\ }\textbf {\bibinfo
  {volume} {18}},\ \bibinfo {pages} {013015} (\bibinfo {year} {2016})}\BibitemShut {NoStop}%
\bibitem [{\citenamefont {Wickenbrock}\ \emph {et~al.}(2016)\citenamefont
  {Wickenbrock}, \citenamefont {Zheng}, \citenamefont {Bougas}, \citenamefont
  {Leefer}, \citenamefont {Afach}, \citenamefont {Jarmola}, \citenamefont
  {Acosta},\ and\ \citenamefont {Budker}}]{wickenbrock_microwave-free_2016}%
  \BibitemOpen
  \bibfield  {author} {\bibinfo {author} {\bibfnamefont {A.}~\bibnamefont
  {Wickenbrock}}, \bibinfo {author} {\bibfnamefont {H.}~\bibnamefont {Zheng}},
  \bibinfo {author} {\bibfnamefont {L.}~\bibnamefont {Bougas}}, \bibinfo
  {author} {\bibfnamefont {N.}~\bibnamefont {Leefer}}, \bibinfo {author}
  {\bibfnamefont {S.}~\bibnamefont {Afach}}, \bibinfo {author} {\bibfnamefont
  {A.}~\bibnamefont {Jarmola}}, \bibinfo {author} {\bibfnamefont {V.~M.}\
  \bibnamefont {Acosta}},\ and\ \bibinfo {author} {\bibfnamefont
  {D.}~\bibnamefont {Budker}},\ }\href {https://doi.org/10.1063/1.4960171} {\bibfield  {journal} {\bibinfo
  {journal} {Appl. Phys. Lett.}\ }\textbf {\bibinfo {volume} {109}},\ \bibinfo
  {pages} {053505} (\bibinfo {year} {2016})}\BibitemShut {NoStop}%
\bibitem [{\citenamefont {Wood}\ \emph
  {et~al.}(2018{\natexlab{a}})\citenamefont {Wood}, \citenamefont {Aeppli},
  \citenamefont {Lilette}, \citenamefont {Fein}, \citenamefont {Stacey},
  \citenamefont {Hollenberg}, \citenamefont {Scholten},\ and\ \citenamefont
  {Martin}}]{wood_t_2-limited_2018}%
  \BibitemOpen
  \bibfield  {author} {\bibinfo {author} {\bibfnamefont {A.~A.}\ \bibnamefont
  {Wood}}, \bibinfo {author} {\bibfnamefont {A.~G.}\ \bibnamefont {Aeppli}},
  \bibinfo {author} {\bibfnamefont {E.}~\bibnamefont {Lilette}}, \bibinfo
  {author} {\bibfnamefont {Y.~Y.}\ \bibnamefont {Fein}}, \bibinfo {author}
  {\bibfnamefont {A.}~\bibnamefont {Stacey}}, \bibinfo {author} {\bibfnamefont
  {L.~C.~L.}\ \bibnamefont {Hollenberg}}, \bibinfo {author} {\bibfnamefont
  {R.~E.}\ \bibnamefont {Scholten}},\ and\ \bibinfo {author} {\bibfnamefont
  {A.~M.}\ \bibnamefont {Martin}},\ }\href
  {https://doi.org/10.1103/PhysRevB.98.174114} {\bibfield  {journal} {\bibinfo
  {journal} {Phys. Rev. B}\ }\textbf {\bibinfo {volume} {98}},\ \bibinfo
  {pages} {174114} (\bibinfo {year} {2018}{\natexlab{a}})}\BibitemShut {NoStop}%
\bibitem [{\citenamefont {Wood}\ \emph
  {et~al.}(2018{\natexlab{b}})\citenamefont {Wood}, \citenamefont {Lilette},
  \citenamefont {Fein}, \citenamefont {Tomek}, \citenamefont {McGuinness},
  \citenamefont {Hollenberg}, \citenamefont {Scholten},\ and\ \citenamefont
  {Martin}}]{wood_quantum_2018}%
  \BibitemOpen
  \bibfield  {author} {\bibinfo {author} {\bibfnamefont {A.~A.}\ \bibnamefont
  {Wood}}, \bibinfo {author} {\bibfnamefont {E.}~\bibnamefont {Lilette}},
  \bibinfo {author} {\bibfnamefont {Y.~Y.}\ \bibnamefont {Fein}}, \bibinfo
  {author} {\bibfnamefont {N.}~\bibnamefont {Tomek}}, \bibinfo {author}
  {\bibfnamefont {L.~P.}\ \bibnamefont {McGuinness}}, \bibinfo {author}
  {\bibfnamefont {L.~C.~L.}\ \bibnamefont {Hollenberg}}, \bibinfo {author}
  {\bibfnamefont {R.~E.}\ \bibnamefont {Scholten}},\ and\ \bibinfo {author}
  {\bibfnamefont {A.~M.}\ \bibnamefont {Martin}},\ }\href
  {https://doi.org/10.1126/sciadv.aar7691} {\bibfield  {journal} {\bibinfo
  {journal} {Science Advances}\ }\textbf {\bibinfo {volume} {4}},\ \bibinfo
  {pages} {eaar7691} (\bibinfo {year} {2018}{\natexlab{b}})}\BibitemShut {NoStop}%
\bibitem [{\citenamefont {Wood}\ \emph {et~al.}(2021)\citenamefont {Wood},
  \citenamefont {Goldblatt}, \citenamefont {Anderson}, \citenamefont
  {Hollenberg}, \citenamefont {Scholten},\ and\ \citenamefont
  {Martin}}]{wood_anisotropic_2021}%
  \BibitemOpen
  \bibfield  {author} {\bibinfo {author} {\bibfnamefont {A.~A.}\ \bibnamefont
  {Wood}}, \bibinfo {author} {\bibfnamefont {R.~M.}\ \bibnamefont {Goldblatt}},
  \bibinfo {author} {\bibfnamefont {R.~P.}\ \bibnamefont {Anderson}}, \bibinfo
  {author} {\bibfnamefont {L.~C.~L.}\ \bibnamefont {Hollenberg}}, \bibinfo
  {author} {\bibfnamefont {R.~E.}\ \bibnamefont {Scholten}},\ and\ \bibinfo
  {author} {\bibfnamefont {A.~M.}\ \bibnamefont {Martin}},\ }\href {http://arxiv.org/abs/2105.07365} {\bibfield
  {journal} {\bibinfo  {journal} {Phys. Rev. B \textbf{104} 085419} (\bibinfo
  {year} {2021})}}\BibitemShut {NoStop}%
\bibitem [{\citenamefont {Wolf}\ \emph {et~al.}(2015)\citenamefont {Wolf},
  \citenamefont {Neumann}, \citenamefont {Nakamura}, \citenamefont {Sumiya},
  \citenamefont {Ohshima}, \citenamefont {Isoya},\ and\ \citenamefont
  {Wrachtrup}}]{wolf_subpicotesla_2015}%
  \BibitemOpen
  \bibfield  {author} {\bibinfo {author} {\bibfnamefont {T.}~\bibnamefont
  {Wolf}}, \bibinfo {author} {\bibfnamefont {P.}~\bibnamefont {Neumann}},
  \bibinfo {author} {\bibfnamefont {K.}~\bibnamefont {Nakamura}}, \bibinfo
  {author} {\bibfnamefont {H.}~\bibnamefont {Sumiya}}, \bibinfo {author}
  {\bibfnamefont {T.}~\bibnamefont {Ohshima}}, \bibinfo {author} {\bibfnamefont
  {J.}~\bibnamefont {Isoya}},\ and\ \bibinfo {author} {\bibfnamefont
  {J.}~\bibnamefont {Wrachtrup}},\ }\href
  {https://doi.org/10.1103/PhysRevX.5.041001} {\bibfield  {journal} {\bibinfo
  {journal} {Phys. Rev. X}\ }\textbf {\bibinfo {volume} {5}},\ \bibinfo {pages}
  {041001} (\bibinfo {year} {2015})}\BibitemShut
  {NoStop}%
\bibitem [{Note1()}]{Note1}%
  \BibitemOpen
  \bibinfo {note} {See Supplementary Material}\BibitemShut {NoStop}%
\bibitem [{Note2()}]{Note2}%
  \BibitemOpen
  \bibinfo {note} {Derived in full in the Supplementary Material}\BibitemShut {NoStop}%
\bibitem [{Note3()}]{Note3}%
  \BibitemOpen
  \bibinfo {note} {A comprehensive description of Allan deviation as it applies to our work is provided in the Supplementary Material.}\BibitemShut {NoStop}%
\bibitem [{\citenamefont {Degen}\ \emph {et~al.}(2017)\citenamefont {Degen},
  \citenamefont {Reinhard},\ and\ \citenamefont
  {Cappellaro}}]{degen_quantum_2017}%
  \BibitemOpen
  \bibfield  {author} {\bibinfo {author} {\bibfnamefont {C.}~\bibnamefont
  {Degen}}, \bibinfo {author} {\bibfnamefont {F.}~\bibnamefont {Reinhard}},\
  and\ \bibinfo {author} {\bibfnamefont {P.}~\bibnamefont {Cappellaro}},\
  }\href
  {https://doi.org/10.1103/RevModPhys.89.035002} {\bibfield  {journal}
  {\bibinfo  {journal} {Rev. Mod. Phys.}\ }\textbf {\bibinfo {volume} {89}},\
  \bibinfo {pages} {035002} (\bibinfo {year} {2017})}\BibitemShut {NoStop}%
\bibitem [{\citenamefont {Zhao}\ \emph {et~al.}(2012)\citenamefont {Zhao},
  \citenamefont {Ho},\ and\ \citenamefont {Liu}}]{zhao_decoherence_2012}%
  \BibitemOpen
  \bibfield  {author} {\bibinfo {author} {\bibfnamefont {N.}~\bibnamefont
  {Zhao}}, \bibinfo {author} {\bibfnamefont {S.-W.}\ \bibnamefont {Ho}},\ and\
  \bibinfo {author} {\bibfnamefont {R.-B.}\ \bibnamefont {Liu}},\ }\href
  {https://doi.org/10.1103/PhysRevB.85.115303} {\bibfield  {journal} {\bibinfo
  {journal} {Phys. Rev. B}\ }\textbf {\bibinfo {volume} {85}},\ \bibinfo
  {pages} {115303} (\bibinfo {year} {2012})}\BibitemShut {NoStop}%
\bibitem [{\citenamefont {Hall}\ \emph {et~al.}(2014)\citenamefont {Hall},
  \citenamefont {Cole},\ and\ \citenamefont {Hollenberg}}]{hall_analytic_2014}%
  \BibitemOpen
  \bibfield  {author} {\bibinfo {author} {\bibfnamefont {L.~T.}\ \bibnamefont
  {Hall}}, \bibinfo {author} {\bibfnamefont {J.~H.}\ \bibnamefont {Cole}},\
  and\ \bibinfo {author} {\bibfnamefont {L.~C.~L.}\ \bibnamefont
  {Hollenberg}},\ }\href
  {https://doi.org/10.1103/PhysRevB.90.075201} {\bibfield  {journal} {\bibinfo
  {journal} {Phys. Rev. B}\ }\textbf {\bibinfo {volume} {90}},\ \bibinfo
  {pages} {075201} (\bibinfo {year} {2014})}\BibitemShut {NoStop}%
\bibitem [{\citenamefont {Stanwix}\ \emph {et~al.}(2010)\citenamefont
  {Stanwix}, \citenamefont {Pham}, \citenamefont {Maze}, \citenamefont
  {Le~Sage}, \citenamefont {Yeung}, \citenamefont {Cappellaro}, \citenamefont
  {Hemmer}, \citenamefont {Yacoby}, \citenamefont {Lukin},\ and\ \citenamefont
  {Walsworth}}]{stanwix_coherence_2010}%
  \BibitemOpen
  \bibfield  {author} {\bibinfo {author} {\bibfnamefont {P.~L.}\ \bibnamefont
  {Stanwix}}, \bibinfo {author} {\bibfnamefont {L.~M.}\ \bibnamefont {Pham}},
  \bibinfo {author} {\bibfnamefont {J.~R.}\ \bibnamefont {Maze}}, \bibinfo
  {author} {\bibfnamefont {D.}~\bibnamefont {Le~Sage}}, \bibinfo {author}
  {\bibfnamefont {T.~K.}\ \bibnamefont {Yeung}}, \bibinfo {author}
  {\bibfnamefont {P.}~\bibnamefont {Cappellaro}}, \bibinfo {author}
  {\bibfnamefont {P.~R.}\ \bibnamefont {Hemmer}}, \bibinfo {author}
  {\bibfnamefont {A.}~\bibnamefont {Yacoby}}, \bibinfo {author} {\bibfnamefont
  {M.~D.}\ \bibnamefont {Lukin}},\ and\ \bibinfo {author} {\bibfnamefont
  {R.~L.}\ \bibnamefont {Walsworth}},\ }\href
  {https://doi.org/10.1103/PhysRevB.82.201201} {\bibfield  {journal} {\bibinfo
  {journal} {Phys. Rev. B}\ }\textbf {\bibinfo {volume} {82}},\ \bibinfo
  {pages} {201201} (\bibinfo {year} {2010})}\BibitemShut {NoStop}%
\bibitem [{\citenamefont {Wood}\ \emph {et~al.}(2017)\citenamefont {Wood},
  \citenamefont {Lilette}, \citenamefont {Fein}, \citenamefont {Perunicic},
  \citenamefont {Hollenberg}, \citenamefont {Scholten},\ and\ \citenamefont
  {Martin}}]{wood_magnetic_2017}%
  \BibitemOpen
  \bibfield  {author} {\bibinfo {author} {\bibfnamefont {A.~A.}\ \bibnamefont
  {Wood}}, \bibinfo {author} {\bibfnamefont {E.}~\bibnamefont {Lilette}},
  \bibinfo {author} {\bibfnamefont {Y.~Y.}\ \bibnamefont {Fein}}, \bibinfo
  {author} {\bibfnamefont {V.~S.}\ \bibnamefont {Perunicic}}, \bibinfo {author}
  {\bibfnamefont {L.~C.~L.}\ \bibnamefont {Hollenberg}}, \bibinfo {author}
  {\bibfnamefont {R.~E.}\ \bibnamefont {Scholten}},\ and\ \bibinfo {author}
  {\bibfnamefont {A.~M.}\ \bibnamefont {Martin}},\ }\href {https://doi.org/10.1038/nphys4221}
  {\bibfield  {journal} {\bibinfo  {journal} {Nature Physics}\ }\textbf
  {\bibinfo {volume} {13}},\ \bibinfo {pages} {nphys4221} (\bibinfo {year}
  {2017})}\BibitemShut {NoStop}%
\bibitem [{Note4()}]{Note4}%
  \BibitemOpen
  \bibinfo {note} {Discussed in greater detail in the Supplementary
  Material}\BibitemShut {NoStop}%
\bibitem [{\citenamefont {Ajoy}\ \emph {et~al.}(2016)\citenamefont {Ajoy},
  \citenamefont {Liu},\ and\ \citenamefont {Cappellaro}}]{ajoy_DC_2016}%
  \BibitemOpen
  \bibfield  {author} {\bibinfo {author} {\bibfnamefont {A.}~\bibnamefont
  {Ajoy}}, \bibinfo {author} {\bibfnamefont {Y.~X.}\ \bibnamefont {Liu}},\ and\
  \bibinfo {author} {\bibfnamefont {P.}~\bibnamefont {Cappellaro}},\ }\href {http://arxiv.org/abs/1611.04691} {\bibfield  {journal} {\bibinfo
  {journal} {arXiv:1611.04691}\ }
  (\bibinfo {year} {2016})}\BibitemShut{NoStop}%
\bibitem [{\citenamefont {Schledorn}\ \emph {et~al.}(2020)\citenamefont
  {Schledorn}, \citenamefont {Malär}, \citenamefont {Torosyan}, \citenamefont
  {Penzel}, \citenamefont {Klose}, \citenamefont {Oss}, \citenamefont {Org},
  \citenamefont {Wang}, \citenamefont {Lecoq}, \citenamefont {Cadalbert},
  \citenamefont {Samoson}, \citenamefont {Böckmann},\ and\ \citenamefont
  {Meier}}]{schledorn_protein_2020}%
  \BibitemOpen
  \bibfield  {author} {\bibinfo {author} {\bibfnamefont {M.}~\bibnamefont
  {Schledorn}}, \bibinfo {author} {\bibfnamefont {A.~A.}\ \bibnamefont
  {Malär}}, \bibinfo {author} {\bibfnamefont {A.}~\bibnamefont {Torosyan}},
  \bibinfo {author} {\bibfnamefont {S.}~\bibnamefont {Penzel}}, \bibinfo
  {author} {\bibfnamefont {D.}~\bibnamefont {Klose}}, \bibinfo {author}
  {\bibfnamefont {A.}~\bibnamefont {Oss}}, \bibinfo {author} {\bibfnamefont
  {M.-L.}\ \bibnamefont {Org}}, \bibinfo {author} {\bibfnamefont
  {S.}~\bibnamefont {Wang}}, \bibinfo {author} {\bibfnamefont {L.}~\bibnamefont
  {Lecoq}}, \bibinfo {author} {\bibfnamefont {R.}~\bibnamefont {Cadalbert}},
  \bibinfo {author} {\bibfnamefont {A.}~\bibnamefont {Samoson}}, \bibinfo
  {author} {\bibfnamefont {A.}~\bibnamefont {Böckmann}},\ and\ \bibinfo
  {author} {\bibfnamefont {B.~H.}\ \bibnamefont {Meier}},\ }\href
  {https://doi.org/10.1002/cbic.202000341} {\bibfield  {journal} {\bibinfo
  {journal} {Chembiochem}\ }\textbf {\bibinfo {volume} {21}},\ \bibinfo {pages}
  {2540} (\bibinfo {year} {2020})}\BibitemShut {NoStop}%
\bibitem [{\citenamefont {Reimann}\ \emph {et~al.}(2018)\citenamefont
  {Reimann}, \citenamefont {Doderer}, \citenamefont {Hebestreit}, \citenamefont
  {Diehl}, \citenamefont {Frimmer}, \citenamefont {Windey}, \citenamefont
  {Tebbenjohanns},\ and\ \citenamefont {Novotny}}]{reimann_GHz_2018}%
  \BibitemOpen
  \bibfield  {author} {\bibinfo {author} {\bibfnamefont {R.}~\bibnamefont
  {Reimann}}, \bibinfo {author} {\bibfnamefont {M.}~\bibnamefont {Doderer}},
  \bibinfo {author} {\bibfnamefont {E.}~\bibnamefont {Hebestreit}}, \bibinfo
  {author} {\bibfnamefont {R.}~\bibnamefont {Diehl}}, \bibinfo {author}
  {\bibfnamefont {M.}~\bibnamefont {Frimmer}}, \bibinfo {author} {\bibfnamefont
  {D.}~\bibnamefont {Windey}}, \bibinfo {author} {\bibfnamefont
  {F.}~\bibnamefont {Tebbenjohanns}},\ and\ \bibinfo {author} {\bibfnamefont
  {L.}~\bibnamefont {Novotny}},\ }\href
  {https://doi.org/10.1103/PhysRevLett.121.033602} {\bibfield  {journal}
  {\bibinfo  {journal} {Phys. Rev. Lett.}\ }\textbf {\bibinfo {volume} {121}},\
  \bibinfo {pages} {033602} (\bibinfo {year} {2018})}\BibitemShut {NoStop}%
\bibitem [{\citenamefont {Ahn}\ \emph {et~al.}(2020)\citenamefont {Ahn},
  \citenamefont {Xu}, \citenamefont {Bang}, \citenamefont {Ju}, \citenamefont
  {Gao},\ and\ \citenamefont {Li}}]{ahn_ultrasensitive_2020}%
  \BibitemOpen
  \bibfield  {author} {\bibinfo {author} {\bibfnamefont {J.}~\bibnamefont
  {Ahn}}, \bibinfo {author} {\bibfnamefont {Z.}~\bibnamefont {Xu}}, \bibinfo
  {author} {\bibfnamefont {J.}~\bibnamefont {Bang}}, \bibinfo {author}
  {\bibfnamefont {P.}~\bibnamefont {Ju}}, \bibinfo {author} {\bibfnamefont
  {X.}~\bibnamefont {Gao}},\ and\ \bibinfo {author} {\bibfnamefont
  {T.}~\bibnamefont {Li}},\ }\href {https://doi.org/10.1038/s41565-019-0605-9} {\bibfield
  {journal} {\bibinfo  {journal} {Nat. Nanotechnol.}\ }\textbf {\bibinfo
  {volume} {15}},\ \bibinfo {pages} {89} (\bibinfo {year} {2020})}\BibitemShut {NoStop}%
\bibitem [{\citenamefont {Csencsics}\ \emph {et~al.}(2019)\citenamefont
  {Csencsics}, \citenamefont {Sitz},\ and\ \citenamefont
  {Schitter}}]{csencsics_fast_2019}%
  \BibitemOpen
  \bibfield  {author} {\bibinfo {author} {\bibfnamefont {E.}~\bibnamefont
  {Csencsics}}, \bibinfo {author} {\bibfnamefont {B.}~\bibnamefont {Sitz}},\
  and\ \bibinfo {author} {\bibfnamefont {G.}~\bibnamefont {Schitter}},\
  }\href
  {https://doi.org/10.1016/j.ifacol.2019.11.689} {\bibfield  {journal}
  {\bibinfo  {journal} {IFAC-PapersOnLine}\ }\bibinfo {series} {8th {IFAC}
  {Symposium} on {Mechatronic} {Systems} {MECHATRONICS} 2019},\ \textbf
  {\bibinfo {volume} {52}},\ \bibinfo {pages} {289} (\bibinfo {year}
  {2019})}\BibitemShut {NoStop}%
\bibitem [{\citenamefont {Perdriat}\ \emph {et~al.}(2021)\citenamefont
  {Perdriat}, \citenamefont {Pellet-Mary}, \citenamefont {Huillery},
  \citenamefont {Rondin},\ and\ \citenamefont
  {Hétet}}]{perdriat_spin-mechanics_2021}%
  \BibitemOpen
  \bibfield  {author} {\bibinfo {author} {\bibfnamefont {M.}~\bibnamefont
  {Perdriat}}, \bibinfo {author} {\bibfnamefont {C.}~\bibnamefont
  {Pellet-Mary}}, \bibinfo {author} {\bibfnamefont {P.}~\bibnamefont
  {Huillery}}, \bibinfo {author} {\bibfnamefont {L.}~\bibnamefont {Rondin}},\
  and\ \bibinfo {author} {\bibfnamefont {G.}~\bibnamefont {Hétet}},\
  }\href
  {https://doi.org/10.3390/mi12060651} {\bibfield  {journal} {\bibinfo
  {journal} {Micromachines}\ }\textbf {\bibinfo {volume} {12}},\ \bibinfo
  {pages} {651} (\bibinfo {year} {2021})}\BibitemShut {NoStop}%
\bibitem [{\citenamefont {Huxter}\ \emph {et~al.}(2022)\citenamefont {Huxter},
  \citenamefont {Palm}, \citenamefont {Davis}, \citenamefont {Welter},
  \citenamefont {Lambert}, \citenamefont {Trassin},\ and\ \citenamefont
  {Degen}}]{huxter_scanning_2022}%
  \BibitemOpen
  \bibfield  {author} {\bibinfo {author} {\bibfnamefont {W.~S.}\ \bibnamefont
  {Huxter}}, \bibinfo {author} {\bibfnamefont {M.~L.}\ \bibnamefont {Palm}},
  \bibinfo {author} {\bibfnamefont {M.~L.}\ \bibnamefont {Davis}}, \bibinfo
  {author} {\bibfnamefont {P.}~\bibnamefont {Welter}}, \bibinfo {author}
  {\bibfnamefont {C.-H.}\ \bibnamefont {Lambert}}, \bibinfo {author}
  {\bibfnamefont {M.}~\bibnamefont {Trassin}},\ and\ \bibinfo {author}
  {\bibfnamefont {C.~L.}\ \bibnamefont {Degen}},\ }\href {http://arxiv.org/abs/2202.09130} {\bibfield  {journal} {\bibinfo
  {journal} {arXiv:2202.09130}} (\bibinfo {year}
  {2022})}\BibitemShut {NoStop}%
\bibitem [{\citenamefont {Boto}\ \emph {et~al.}(2018)\citenamefont {Boto},
  \citenamefont {Holmes}, \citenamefont {Leggett}, \citenamefont {Roberts},
  \citenamefont {Shah}, \citenamefont {Meyer}, \citenamefont {Muñoz},
  \citenamefont {Mullinger}, \citenamefont {Tierney}, \citenamefont {Bestmann},
  \citenamefont {Barnes}, \citenamefont {Bowtell},\ and\ \citenamefont
  {Brookes}}]{boto_moving_2018}%
  \BibitemOpen
  \bibfield  {author} {\bibinfo {author} {\bibfnamefont {E.}~\bibnamefont
  {Boto}}, \bibinfo {author} {\bibfnamefont {N.}~\bibnamefont {Holmes}},
  \bibinfo {author} {\bibfnamefont {J.}~\bibnamefont {Leggett}}, \bibinfo
  {author} {\bibfnamefont {G.}~\bibnamefont {Roberts}}, \bibinfo {author}
  {\bibfnamefont {V.}~\bibnamefont {Shah}}, \bibinfo {author} {\bibfnamefont
  {S.~S.}\ \bibnamefont {Meyer}}, \bibinfo {author} {\bibfnamefont {L.~D.}\
  \bibnamefont {Muñoz}}, \bibinfo {author} {\bibfnamefont {K.~J.}\
  \bibnamefont {Mullinger}}, \bibinfo {author} {\bibfnamefont {T.~M.}\
  \bibnamefont {Tierney}}, \bibinfo {author} {\bibfnamefont {S.}~\bibnamefont
  {Bestmann}}, \bibinfo {author} {\bibfnamefont {G.~R.}\ \bibnamefont
  {Barnes}}, \bibinfo {author} {\bibfnamefont {R.}~\bibnamefont {Bowtell}},\
  and\ \bibinfo {author} {\bibfnamefont {M.~J.}\ \bibnamefont {Brookes}},\
  }\href {https://doi.org/10.1038/nature26147} {\bibfield  {journal}
  {\bibinfo  {journal} {Nature}\ }\textbf {\bibinfo {volume} {555}},\ \bibinfo
  {pages} {657} (\bibinfo {year} {2018})}\BibitemShut {NoStop}%
\end{thebibliography}

\begin{thebibliography}{9}%

\makeatletter
\providecommand \@ifxundefined [1]{%
 \@ifx{#1\undefined}
}%
\providecommand \@ifnum [1]{%
 \ifnum #1\expandafter \@firstoftwo
 \else \expandafter \@secondoftwo
 \fi
}%
\providecommand \@ifx [1]{%
 \ifx #1\expandafter \@firstoftwo
 \else \expandafter \@secondoftwo
 \fi
}%
\providecommand \natexlab [1]{#1}%
\providecommand \enquote  [1]{``#1''}%
\providecommand \bibnamefont  [1]{#1}%
\providecommand \bibfnamefont [1]{#1}%
\providecommand \citenamefont [1]{#1}%
\providecommand \href@noop [0]{\@secondoftwo}%
\providecommand \href [0]{\begingroup \@sanitize@url \@href}%
\providecommand \@href[1]{\@@startlink{#1}\@@href}%
\providecommand \@@href[1]{\endgroup#1\@@endlink}%
\providecommand \@sanitize@url [0]{\catcode `\\12\catcode `\$12\catcode
  `\&12\catcode `\#12\catcode `\^12\catcode `\_12\catcode `\%12\relax}%
\providecommand \@@startlink[1]{}%
\providecommand \@@endlink[0]{}%
\providecommand \url  [0]{\begingroup\@sanitize@url \@url }%
\providecommand \@url [1]{\endgroup\@href {#1}{\urlprefix }}%
\providecommand \urlprefix  [0]{URL }%
\providecommand \Eprint [0]{\href }%
\providecommand \doibase [0]{https://doi.org/}%
\providecommand \selectlanguage [0]{\@gobble}%
\providecommand \bibinfo  [0]{\@secondoftwo}%
\providecommand \bibfield  [0]{\@secondoftwo}%
\providecommand \translation [1]{[#1]}%
\providecommand \BibitemOpen [0]{}%
\providecommand \bibitemStop [0]{}%
\providecommand \bibitemNoStop [0]{.\EOS\space}%
\providecommand \EOS [0]{\spacefactor3000\relax}%
\providecommand \BibitemShut  [1]{\csname bibitem#1\endcsname}%
\let\auto@bib@innerbib\@empty
\bibitem [{\citenamefont {Wood}\ \emph {et~al.}(2018)\citenamefont {Wood},
  \citenamefont {Aeppli}, \citenamefont {Lilette}, \citenamefont {Fein},
  \citenamefont {Stacey}, \citenamefont {Hollenberg}, \citenamefont
  {Scholten},\ and\ \citenamefont {Martin}}]{wood_t_2-limited_2018s}%
  \BibitemOpen
  \bibfield  {author} {\bibinfo {author} {\bibfnamefont {A.~A.}\ \bibnamefont
  {Wood}}, \bibinfo {author} {\bibfnamefont {A.~G.}\ \bibnamefont {Aeppli}},
  \bibinfo {author} {\bibfnamefont {E.}~\bibnamefont {Lilette}}, \bibinfo
  {author} {\bibfnamefont {Y.~Y.}\ \bibnamefont {Fein}}, \bibinfo {author}
  {\bibfnamefont {A.}~\bibnamefont {Stacey}}, \bibinfo {author} {\bibfnamefont
  {L.~C.~L.}\ \bibnamefont {Hollenberg}}, \bibinfo {author} {\bibfnamefont
  {R.~E.}\ \bibnamefont {Scholten}},\ and\ \bibinfo {author} {\bibfnamefont
  {A.~M.}\ \bibnamefont {Martin}},\ }\href
  {https://doi.org/10.1103/PhysRevB.98.174114} {\bibfield  {journal} {\bibinfo
  {journal} {Phys. Rev. B}\ }\textbf {\bibinfo {volume} {98}},\ \bibinfo
  {pages} {174114} (\bibinfo {year} {2018})}\BibitemShut {NoStop}%
\bibitem [{\citenamefont {Binder}\ \emph {et~al.}(2017)\citenamefont {Binder},
  \citenamefont {Stark}, \citenamefont {Tomek}, \citenamefont {Scheuer},
  \citenamefont {Frank}, \citenamefont {Jahnke}, \citenamefont {Müller},
  \citenamefont {Schmitt}, \citenamefont {Metsch}, \citenamefont {Unden},
  \citenamefont {Gehring}, \citenamefont {Huck}, \citenamefont {Andersen},
  \citenamefont {Rogers},\ and\ \citenamefont {Jelezko}}]{binder_qudi_2017s}%
  \BibitemOpen
  \bibfield  {author} {\bibinfo {author} {\bibfnamefont {J.~M.}\ \bibnamefont
  {Binder}}, \bibinfo {author} {\bibfnamefont {A.}~\bibnamefont {Stark}},
  \bibinfo {author} {\bibfnamefont {N.}~\bibnamefont {Tomek}}, \bibinfo
  {author} {\bibfnamefont {J.}~\bibnamefont {Scheuer}}, \bibinfo {author}
  {\bibfnamefont {F.}~\bibnamefont {Frank}}, \bibinfo {author} {\bibfnamefont
  {K.~D.}\ \bibnamefont {Jahnke}}, \bibinfo {author} {\bibfnamefont
  {C.}~\bibnamefont {Müller}}, \bibinfo {author} {\bibfnamefont
  {S.}~\bibnamefont {Schmitt}}, \bibinfo {author} {\bibfnamefont {M.~H.}\
  \bibnamefont {Metsch}}, \bibinfo {author} {\bibfnamefont {T.}~\bibnamefont
  {Unden}}, \bibinfo {author} {\bibfnamefont {T.}~\bibnamefont {Gehring}},
  \bibinfo {author} {\bibfnamefont {A.}~\bibnamefont {Huck}}, \bibinfo {author}
  {\bibfnamefont {U.~L.}\ \bibnamefont {Andersen}}, \bibinfo {author}
  {\bibfnamefont {L.~J.}\ \bibnamefont {Rogers}},\ and\ \bibinfo {author}
  {\bibfnamefont {F.}~\bibnamefont {Jelezko}},\ }\href
  {https://doi.org/10.1016/j.softx.2017.02.001} {\bibfield  {journal} {\bibinfo
   {journal} {SoftwareX}\ }\textbf {\bibinfo {volume} {6}},\ \bibinfo {pages}
  {85} (\bibinfo {year} {2017})}\BibitemShut {NoStop}%
\bibitem [{\citenamefont {Bauch}\ \emph {et~al.}(2018)\citenamefont {Bauch},
  \citenamefont {Hart}, \citenamefont {Schloss}, \citenamefont {Turner},
  \citenamefont {Barry}, \citenamefont {Kehayias}, \citenamefont {Singh},\ and\
  \citenamefont {Walsworth}}]{bauch_ultralong_2018s}%
  \BibitemOpen
  \bibfield  {author} {\bibinfo {author} {\bibfnamefont {E.}~\bibnamefont
  {Bauch}}, \bibinfo {author} {\bibfnamefont {C.~A.}\ \bibnamefont {Hart}},
  \bibinfo {author} {\bibfnamefont {J.~M.}\ \bibnamefont {Schloss}}, \bibinfo
  {author} {\bibfnamefont {M.~J.}\ \bibnamefont {Turner}}, \bibinfo {author}
  {\bibfnamefont {J.~F.}\ \bibnamefont {Barry}}, \bibinfo {author}
  {\bibfnamefont {P.}~\bibnamefont {Kehayias}}, \bibinfo {author}
  {\bibfnamefont {S.}~\bibnamefont {Singh}},\ and\ \bibinfo {author}
  {\bibfnamefont {R.~L.}\ \bibnamefont {Walsworth}},\ }\href
  {https://doi.org/10.1103/PhysRevX.8.031025} {\bibfield  {journal} {\bibinfo
  {journal} {Phys. Rev. X}\ }\textbf {\bibinfo {volume} {8}},\ \bibinfo {pages}
  {031025} (\bibinfo {year} {2018})}\BibitemShut
  {NoStop}%
\bibitem [{\citenamefont {Wolf}\ \emph {et~al.}(2015)\citenamefont {Wolf},
  \citenamefont {Neumann}, \citenamefont {Nakamura}, \citenamefont {Sumiya},
  \citenamefont {Ohshima}, \citenamefont {Isoya},\ and\ \citenamefont
  {Wrachtrup}}]{wolf_subpicotesla_2015s}%
  \BibitemOpen
  \bibfield  {author} {\bibinfo {author} {\bibfnamefont {T.}~\bibnamefont
  {Wolf}}, \bibinfo {author} {\bibfnamefont {P.}~\bibnamefont {Neumann}},
  \bibinfo {author} {\bibfnamefont {K.}~\bibnamefont {Nakamura}}, \bibinfo
  {author} {\bibfnamefont {H.}~\bibnamefont {Sumiya}}, \bibinfo {author}
  {\bibfnamefont {T.}~\bibnamefont {Ohshima}}, \bibinfo {author} {\bibfnamefont
  {J.}~\bibnamefont {Isoya}},\ and\ \bibinfo {author} {\bibfnamefont
  {J.}~\bibnamefont {Wrachtrup}},\ }\href
  {https://doi.org/10.1103/PhysRevX.5.041001} {\bibfield  {journal} {\bibinfo
  {journal} {Phys. Rev. X}\ }\textbf {\bibinfo {volume} {5}},\ \bibinfo {pages}
  {041001} (\bibinfo {year} {2015})}\BibitemShut
  {NoStop}%
\bibitem [{\citenamefont {Michl}\ \emph {et~al.}(2019)\citenamefont {Michl},
  \citenamefont {Steiner}, \citenamefont {Denisenko}, \citenamefont {Bülau},
  \citenamefont {Zimmermann}, \citenamefont {Nakamura}, \citenamefont {Sumiya},
  \citenamefont {Onoda}, \citenamefont {Neumann}, \citenamefont {Isoya},\ and\
  \citenamefont {Wrachtrup}}]{michl_robust_2019s}%
  \BibitemOpen
  \bibfield  {author} {\bibinfo {author} {\bibfnamefont {J.}~\bibnamefont
  {Michl}}, \bibinfo {author} {\bibfnamefont {J.}~\bibnamefont {Steiner}},
  \bibinfo {author} {\bibfnamefont {A.}~\bibnamefont {Denisenko}}, \bibinfo
  {author} {\bibfnamefont {A.}~\bibnamefont {Bülau}}, \bibinfo {author}
  {\bibfnamefont {A.}~\bibnamefont {Zimmermann}}, \bibinfo {author}
  {\bibfnamefont {K.}~\bibnamefont {Nakamura}}, \bibinfo {author}
  {\bibfnamefont {H.}~\bibnamefont {Sumiya}}, \bibinfo {author} {\bibfnamefont
  {S.}~\bibnamefont {Onoda}}, \bibinfo {author} {\bibfnamefont
  {P.}~\bibnamefont {Neumann}}, \bibinfo {author} {\bibfnamefont
  {J.}~\bibnamefont {Isoya}},\ and\ \bibinfo {author} {\bibfnamefont
  {J.}~\bibnamefont {Wrachtrup}},\ }\href
  {https://doi.org/10.1021/acs.nanolett.9b00900} {\bibfield  {journal}
  {\bibinfo  {journal} {Nano Lett.}\ }\textbf {\bibinfo {volume} {19}},\
  \bibinfo {pages} {4904} (\bibinfo {year} {2019})}\BibitemShut {NoStop}%
\bibitem [{\citenamefont {Fescenko}\ \emph {et~al.}(2020)\citenamefont
  {Fescenko}, \citenamefont {Jarmola}, \citenamefont {Savukov}, \citenamefont
  {Kehayias}, \citenamefont {Smits}, \citenamefont {Damron}, \citenamefont
  {Ristoff}, \citenamefont {Mosavian},\ and\ \citenamefont
  {Acosta}}]{fescenko_diamond_2020s}%
  \BibitemOpen
  \bibfield  {author} {\bibinfo {author} {\bibfnamefont {I.}~\bibnamefont
  {Fescenko}}, \bibinfo {author} {\bibfnamefont {A.}~\bibnamefont {Jarmola}},
  \bibinfo {author} {\bibfnamefont {I.}~\bibnamefont {Savukov}}, \bibinfo
  {author} {\bibfnamefont {P.}~\bibnamefont {Kehayias}}, \bibinfo {author}
  {\bibfnamefont {J.}~\bibnamefont {Smits}}, \bibinfo {author} {\bibfnamefont
  {J.}~\bibnamefont {Damron}}, \bibinfo {author} {\bibfnamefont
  {N.}~\bibnamefont {Ristoff}}, \bibinfo {author} {\bibfnamefont
  {N.}~\bibnamefont {Mosavian}},\ and\ \bibinfo {author} {\bibfnamefont
  {V.~M.}\ \bibnamefont {Acosta}},\ }\href
  {https://doi.org/10.1103/PhysRevResearch.2.023394} {\bibfield  {journal}
  {\bibinfo  {journal} {Phys. Rev. Research}\ }\textbf {\bibinfo {volume}
  {2}},\ \bibinfo {pages} {023394} (\bibinfo {year} {2020})}\BibitemShut {NoStop}%
\bibitem [{\citenamefont {Degen}\ \emph {et~al.}(2017)\citenamefont {Degen},
  \citenamefont {Reinhard},\ and\ \citenamefont
  {Cappellaro}}]{degen_quantum_2017s}%
  \BibitemOpen
  \bibfield  {author} {\bibinfo {author} {\bibfnamefont {C.}~\bibnamefont
  {Degen}}, \bibinfo {author} {\bibfnamefont {F.}~\bibnamefont {Reinhard}},\
  and\ \bibinfo {author} {\bibfnamefont {P.}~\bibnamefont {Cappellaro}},\
  }\href {https://doi.org/10.1103/RevModPhys.89.035002} {\bibfield  {journal}
  {\bibinfo  {journal} {Rev. Mod. Phys.}\ }\textbf {\bibinfo {volume} {89}},\
  \bibinfo {pages} {035002} (\bibinfo {year} {2017})}\BibitemShut {NoStop}%
\bibitem [{\citenamefont {Wood}\ \emph {et~al.}(2017)\citenamefont {Wood},
  \citenamefont {Lilette}, \citenamefont {Fein}, \citenamefont {Perunicic},
  \citenamefont {Hollenberg}, \citenamefont {Scholten},\ and\ \citenamefont
  {Martin}}]{wood_magnetic_2017s}%
  \BibitemOpen
  \bibfield  {author} {\bibinfo {author} {\bibfnamefont {A.~A.}\ \bibnamefont
  {Wood}}, \bibinfo {author} {\bibfnamefont {E.}~\bibnamefont {Lilette}},
  \bibinfo {author} {\bibfnamefont {Y.~Y.}\ \bibnamefont {Fein}}, \bibinfo
  {author} {\bibfnamefont {V.~S.}\ \bibnamefont {Perunicic}}, \bibinfo {author}
  {\bibfnamefont {L.~C.~L.}\ \bibnamefont {Hollenberg}}, \bibinfo {author}
  {\bibfnamefont {R.~E.}\ \bibnamefont {Scholten}},\ and\ \bibinfo {author}
  {\bibfnamefont {A.~M.}\ \bibnamefont {Martin}},\ }\href
  {https://doi.org/10.1038/nphys4221} {\bibfield  {journal} {\bibinfo
  {journal} {Nature Physics}\ }\textbf {\bibinfo {volume} {13}},\ \bibinfo
  {pages} {nphys4221} (\bibinfo {year} {2017})}\BibitemShut {NoStop}%
\bibitem [{\citenamefont {Stanwix}\ \emph {et~al.}(2010)\citenamefont
  {Stanwix}, \citenamefont {Pham}, \citenamefont {Maze}, \citenamefont
  {Le~Sage}, \citenamefont {Yeung}, \citenamefont {Cappellaro}, \citenamefont
  {Hemmer}, \citenamefont {Yacoby}, \citenamefont {Lukin},\ and\ \citenamefont
  {Walsworth}}]{stanwix_coherence_2010s}%
  \BibitemOpen
  \bibfield  {author} {\bibinfo {author} {\bibfnamefont {P.~L.}\ \bibnamefont
  {Stanwix}}, \bibinfo {author} {\bibfnamefont {L.~M.}\ \bibnamefont {Pham}},
  \bibinfo {author} {\bibfnamefont {J.~R.}\ \bibnamefont {Maze}}, \bibinfo
  {author} {\bibfnamefont {D.}~\bibnamefont {Le~Sage}}, \bibinfo {author}
  {\bibfnamefont {T.~K.}\ \bibnamefont {Yeung}}, \bibinfo {author}
  {\bibfnamefont {P.}~\bibnamefont {Cappellaro}}, \bibinfo {author}
  {\bibfnamefont {P.~R.}\ \bibnamefont {Hemmer}}, \bibinfo {author}
  {\bibfnamefont {A.}~\bibnamefont {Yacoby}}, \bibinfo {author} {\bibfnamefont
  {M.~D.}\ \bibnamefont {Lukin}},\ and\ \bibinfo {author} {\bibfnamefont
  {R.~L.}\ \bibnamefont {Walsworth}},\ }\href
  {https://doi.org/10.1103/PhysRevB.82.201201} {\bibfield  {journal} {\bibinfo
  {journal} {Phys. Rev. B}\ }\textbf {\bibinfo {volume} {82}},\ \bibinfo
  {pages} {201201} (\bibinfo {year} {2010})}\BibitemShut {NoStop}%
\end{thebibliography}
\end{document}